\PassOptionsToPackage{pdfpagelabels=false}{hyperref} 
\documentclass[fleqn,usenatbib,useAMS]{mnras}

\usepackage[T1]{fontenc}
\usepackage{ae,aecompl}
\usepackage{times} 
\usepackage{comment}
\usepackage{ulem}
\usepackage{graphicx}	
\usepackage{amsmath}	
\usepackage{amssymb}	
\usepackage{color}
\usepackage{upgreek} 
\usepackage[utf8]{inputenc}
\usepackage{slashed}
\usepackage{mathtools}
\usepackage[outline]{contour}
\usepackage{upgreek}
\usepackage{textcomp} 

\makeatletter
\newcommand{\vast}{\bBigg@{4}}
\newcommand{\Vast}{\bBigg@{5}}

\makeatother

\newcommand{\fett}[1]{\boldsymbol{#1}}

\newcommand{\be}{\begin{equation}}
\newcommand{\ee}{\end{equation}}
\newcommand{\D}{\upartial_a^{\rm L}}
\newcommand{\R}{\mathfrak{R}_{a}}

\contourlength{0.4pt} 
\contournumber{10} 

\title[Quasi-spherical collapse of matter]{Quasi-spherical collapse of matter in \contour{black}{$\Lambda$}CDM}

\author[C.\ Rampf]{Cornelius Rampf\thanks{\!\!Marie Sk\l odowska--Curie Fellow; \,\,e-mail: cornelius.rampf@oca.eu}
   \\
Laboratoire Lagrange, Universit\'e C\^ote d'Azur, Observatoire de la C\^ote d'Azur, CNRS, Blvd de l'Observatoire, CS 34229, 06304 Nice, France \\
Institute for Theoretical Physics (ITP), Philosophenweg 16, University of Heidelberg,  D-69120 Heidelberg, Germany, and \\
Department of Physics, Israel Institute of Technology -- Technion, Haifa 32000, Israel 
}

\date{\today}

\pubyear{2018}

\begin{document}

\label{firstpage}
\pagerange{\pageref{firstpage}--\pageref{lastpage}}
\maketitle

\begin{abstract}
We report the findings of new exact analytical solutions to the cosmological fluid equations, 
namely for the case where the initial conditions are perturbatively close to a spherical top-hat profile.
To do so we enable a fluid description in a Lagrangian-coordinates approach,
and prove the convergence of the Taylor-series representation of the Lagrangian displacement field until the time of collapse (``shell-crossing'').
This allows the determination of the time for quasi-spherical collapse,
which is shown to happen generically earlier than in the spherical case.
For pedagogical reasons,  calculations are first given for a spatially flat universe that 
is only filled with a non-relativistic component of cold dark matter (CDM).
Then, the methodology is updated to a $\Lambda$CDM Universe,  
with the inclusion of a cosmological constant~$\Lambda>0$. 
\end{abstract}


\section{Introduction}
\label{sec:intro}

The spherical collapse model (SCM) is central to many aspects of cosmology. 
Although being a simplified collapse scenario, its practical use can be justified by statistical arguments from peaks theory of \citet{Bardeen:1985tr}, 
which predict that high-density peaks in the Universe tend to be more spherically symmetric than low-density peaks.

Within the context of General Relativity, the non-linear solution of the spherical collapse is a spherically symmetric space-time of a collapsed region occupied by homogeneous matter, given by the Friedmann--Lema\^itre--Robertson--Walker (FLRW) solution with positive spatial curvature. Furthermore, it is well-known that there exists an exact parametric solution for the spherical collapse, at least for simplified cosmologies   (see e.g.\ \citealt{1967ApJ...147..859P,1969PThPh..42....9T,1972ApJ...176....1G,1994ApJ...431..486B}).

Important applications of the SCM include the analytical prediction 
of the shape and position of the baryon acoustic oscillation feature, the latter being
a standard distance scale imprinted into the clustering statistics of tracers within the large-scale structure (LSS), such as galaxies and their host halos (see e.g.\ \citealt{Desjacques:2010gz,Paranjape:2012jt}).
 Another application of the SCM concerns the determination 
of the abundance of primordial black holes (see e.g.\ \citealt{Carr:2016drx}).
In many of these applications -- be it relevant for galaxy, halo or primordial black hole formation, the SCM is employed 
to predict the threshold of linear density fluctuation (at collapse time). This density threshold is then frequently used as the input in phenomenological models that aim to determine the abundance, mass or shape of a given object. 
The SCM is also highly relevant for extracting accurately the cosmological parameters from the probablity distribution functions of spherically averaged densities and velocity divergences, as shown by \citet{Uhlemann:2016wug,Uhlemann:2017hgi}.
The standard framework of the SCM can also be extended/generalized to incorporate the effects of massive neutrino clustering, see \citet{Ichiki:2011ue,LoVerde:2014rxa}.

For many cosmological scenarios,
the collapse can not be modelled by the parametric solution to the dynamics of a closed FLRW universe. Instead one is led to investigate the collapse directly at the level of the equations of motion.
This is for example necessary when incorporating the effect of shear or rotation, as has been done by \citet{Reischke2018}. 
Another interesting example is when investigating the collapse within the framework of
general modifications of the gravitational theory, such as the class of $f(R)$ theories \citep{Starobinsky:2007hu,Hu:2007nk}. There, the appearance of non-local terms in the action of gravity violates the validity of Birkhoff's theorem -- a central requirement to model the matter collapse in terms of a simplistic FLRW model.
See e.g.\ \citet{Borisov:2011fu,Lombriser:2013wta,Kopp:2013lea} for related semi-analytic works.

Apart from the above examples, the SCM is not a suitable framework for incorporating a host of other physical effects. In particular, it is a poor model for {\it realistic matter collapse} which is well-known to be not exactly spherical.
For this reason, 
perhaps one of the most natural advancements to the SCM is the ellipsoidal collapse model, with pioneering works, amongst others, by \citet{1973A&A....27....1I,Bond:1993we}. Based on these works,
\citet{Sheth:1999su} obtained a fitting formula for ellipsoidal collapse which
approximately agrees with results from numerical simulations.

On the theory side, over the past decades 
there have been also efforts to obtain approximative collapse models
that go beyond exact sphericity. The earliest attempts for departing
from spherical symmetry is the work by \citet{Zeldovich:1969sb}, who employed a first-order Lagrangian-coordinates formulation of a matter fluid, leading to the well-known Zel'dovich approximation.
It turns out that such low-order perturbative solutions  do not deliver good approximations for a collapse of initial inhomogeneities that are close to spherical.
Better approximations of such collapse problems have been employed by
\citet{1979ApJ...231....1W,Bond:1993we,Shen:2005wd}, who introduced and refined the
so-called triaxial collapse model. 
There, instead of attempting to solve the underlying fluid equations, the collapse problem is reformulated into an approximative set of three evolution equations for the principal axes of a homogeneous ellipsoid.

In the present paper we solve for the matter collapse analytically by virtue of the cosmological fluid equations.
One important difference between the aforementioned collapse models and the present work is, that we work with initial conditions that represent a perturbed spherical collapse -- the so-called quasi-spherical collapse.
As a consequence of the used methodology, it turns out that the collapse problem comes with a mathematically convergent description, thereby establishing a new class of exact analytical solutions for the cosmological fluid equations. 
Generally, exact analytical solutions to the fluid equations are a very rare-case scenario; so far the only known exact solutions are those for one-dimensional \citep{Novikov:2010ta,Zeldovich:1969sb} and quasi-one-dimensional collapse \citep{Rampf:2017jan}.

The methodology of the present approach is as follows.
We work in a Cartesian coordinate system for the Euler--Poisson equations, and therein employ a Lagrangian-coordinates formulation. 
The last part is crucial since the use of Lagrangian coordinates regularizes the highly singular collapse problem, which is not at all the case in Eulerian coordinates where the density becomes singular.
For the pure spherical case, we use initial conditions that are, to the zeroth order in some expansion parameter~$\epsilon >0$, identical with those that resemble the classical spherical collapse of matter. This way, one obtains an infinite Taylor series for the Lagrangian displacement field, whose low-order Taylor coefficients have been derived by \citet{Munshi:1994zb,Wagner:2015gva}. 
Whether the Taylor series for the displacement is convergent -- even for the case of exact sphericity, however, was not known, since addressing such questions usually requires the knowledge of the limiting behaviour of the Taylor coefficients at arbitrary large orders. 
In this paper, we formally go to all orders in the Taylor series, which allows us to determine the radius of convergence of the Taylor series. 

Then, by going to first order in the small expansion parameter~$\epsilon$, we switch on the arbitrary asymmetric perturbation in the initial conditions, and determine the recursion relations for the Taylor coefficients of the perturbed displacement field.
As a consequence, we obtain a new exact solution to the Euler--Poisson equations, namely for the quasi-spherical matter collapse.

This paper is organized as follows. 
For pedagogical reasons we choose for the beginning parts of the paper (sections~\ref{sec:EPequations}--\ref{sec:pert}) an Einstein-de Sitter (EdS) model, and later update to the commonly accepted cosmological model, the $\Lambda$CDM model (section~\ref{sec:LCDM}). 
In the following section we briefly review the cosmological fluid equations, first in Eulerian and then in Lagrangian coordinates.
In section~\ref{sec:problem} we formulate the perturbation problem, the appropriate initial conditions, and provide the solution {\it Ansatz} in Lagrangian space.
Then, in sections~\ref{sec:unpert} and~\ref{sec:pert} we solve the problem respectively to zeroth order and first order in~$\epsilon$.
Sections~\ref{sec:time} and~\ref{sec:pertfurther}
are devoted to the calculation of the time of perturbed collapse  and the linear density threshold, both relevant as input for e.g.\ halo models.
In section~\ref{sec:LCDM}, we then generalize our results to the $\Lambda$CDM model.
Finally, a discussion and a summary of our results is given in in section~\ref{sec:disc}.

\section{Euler--Poisson equations (E\lowercase{d}S universe)}
\label{sec:EPequations}

\subsection{Basic equations in Eulerian coordinates}\label{sec:Euler}

The cosmological fluid equations can be formulated in comoving coordinates
$\fett{x}= \fett{r}/a$, where $\fett{r}$ is the physical space coordinate and $a$
the cosmic scale factor. The evolution of the latter is given by the usual Friedmann equations. In the present and following sections~\ref{sec:problem}--\ref{sec:pert}, for simplicity,
we choose for the cosmological model an
Einstein--de Sitter (EdS) universe. 
See section~\ref{sec:LCDM} where we generalize our results to the $\Lambda$CDM Universe (where expressions tend to become more cluttered).

In an EdS universe it is assumed that the only evolving energy component is the cold dark matter (CDM); the cosmological constant ($\Lambda$) and the spatial curvature are set to zero. 
The fluid equations for the CDM component are 
\citep{1992STIN...9519341S,1994PhyD...77..342S,Brenier:2003xs}
\begin{subequations} \label{fluidequations}
\begin{align}
 &\upartial_a \fett{v} + (\fett{v} \cdot \nabla) \fett{v} = -\frac{3}{2a} \left( \fett{v} + \nabla \varphi_{\rm g}\right) \,,  \label{eq:Euler} \\
 &\upartial_a \delta + \nabla \cdot \left[ (1+\delta) \fett{v} \right] = 0 \,, \\
 &\nabla^2 \varphi_{\rm g} = \frac \delta a \,, \label{eq:Poisson}
\end{align}
\end{subequations}
where $\fett{v}$ is the peculiar velocity and $\delta  =(\rho - \bar\rho)/\bar \rho$ the density contrast of matter. We make use of the linear growth time $a$, which, for an EdS universe is identical to the cosmic scale factor. As pointed out by \citet{Zheligovsky:2013eca,Rampf:2015mza,Rampf:2017jan}, enabling $a$ as the time variable is essential when investigating the time-analyticity  of the Lagrangian map.

Before considering the Lagrangian-coordinates approach, let us briefly discuss the 
properties of the fluid equations at arbitrary short times. Formally linearizing the three equations~\eqref{fluidequations}, it is straightforward to obtain a single differential equation for the density contrast  (see e.g.\ \citealt{Peebles1980}). This second-order differential equation has two power-law solutions for the density,  one is decaying as $a^{-3/2}$ and the other is growing linearly in $a$. 
From these observations, it becomes evident that the following boundary conditions select the growing-mode and curl-free solution of the fluid equations \citep{Brenier:2003xs},
\be \label{slaving}
 \delta^{\rm (init)} =0\,, \qquad \fett{v}^{\rm (init)} = - \nabla \varphi_{\rm g}^{\rm (init)} \,,
\ee
where ``(init)'' refers to the evaluation at initial time $a=0$. 
Thanks to these {\it slaving conditions}, the solutions of the fluid equations are, for sufficiently early times, time-analytic and thus devoid of any catastrophic behaviour.
Real singularities none the less appear at the instant of shell-crossing,
where particle trajectories intersect for the first time and the density becomes infinite.

\subsection{Basic equations in Lagrangian coordinates}

Let us now turn to the Lagrangian formulation of the fluid equations~\eqref{fluidequations}. 
We denote the Lagrangian coordinates by $\fett{q}$, with components
$q_i$ ($i$=1,2,3). A partial derivative with respect to $q_i$ acting
on a given function $f$ is denoted by $f_{,i}$. Summation over repeated indices is implied.
Let $\fett{q} \mapsto {\fett{x}}(\fett{q},a)$ 
be the Lagrangian map from the initial ($a\!=\!0$)
position $\fett{q}$ to the Eulerian position ${\fett{x}}$ at \mbox{time $a$.}
The map satisfies $\fett{v}(\fett{x}(\fett{q},a),a) = \dot{\fett{x}}(\fett{q},a)$,  
where the overdot is the Lagrangian time derivative (sometimes also denoted with $\upartial_a^{\rm L}$).
At initial time ($a=0$), the velocity is
\be
 \fett{v}^{\rm (init)}(\fett{q}) = \fett{v}(\fett{x}(\fett{q},0),0) \,,
\ee
which agrees with the initial Eulerian velocity.
Mass conservation is, until the first shell-crossing, given by 
\be  \label{lagmass}
 \delta =1/J -1 \,,
\ee 
where $J= \det( x_{i,j} )$ is the Jacobian, which is the determinant of the Jacobian matrix $x_{i,j}$. With these definitions, 
the fluid equations equations can be written in 
Lagrangian coordinates in compact form \citep{Rampf:2017jan},
\begin{subequations} \label{eqs:lag}
\begin{align}
 &\varepsilon_{ikl} \varepsilon_{jmn} \, x_{k,m}\, x_{l,n} \R x_{i,j} = 3 \left( J-1 \right) \,, \label{eq:scalarLag} \\
 &\varepsilon_{ijk} \, \dot x_{l,j} x_{l,k} = 0 \,, \label{eq:Cauchy} 
\end{align}
\end{subequations}
where we have defined the operator $\R \equiv a^2 \left(\upartial_{a}^{\rm L}\right)^2 + (3a/2) \D$, 
and $\varepsilon_{ijk}$ is the fundamental antisymmetric tensor.
Equation~\eqref{eq:scalarLag} is a scalar equation that is obtained by combining equations~\eqref{eq:Euler} and \eqref{eq:Poisson} in Lagrangian coordinates, as well as taking mass conservation~\eqref{lagmass} into account.
Equations~\eqref{eq:Cauchy} are of vectorial character and state the conservation of the zero-vorticity (which holds until shell-crossing) written in Lagrangian coordinates.
 Calculational details about equations~\eqref{eqs:lag} are given by  \citet{Ehlers:1996wg} and  \citet{Zheligovsky:2013eca}.
General derivations of the Lagrangian evolution equations are given by \citet{Buchert:1987xy}, \citet{Rampf:2012xa} and references therein.

\section{Problem and solution Ansatz}\label{sec:problem}

Matter collapse that can be exactly reduced to a spherical problem is degenerate; given the nature of the initial (Gaussian) random density fluctuations, the probability of finding such objects in the LSS is zero. Furthermore, even just a small random perturbation that is added to, say, a spherical overdensity is crucial as it decides shape and orientiation of the collapsed object. 
Initial conditions (ICs) that resemble such a problem are introduced in the following section, and an appropriate solution {\it Ansatz} is given in section~\ref{sec:solAnsatz}.

\subsection{Initial conditions}

In the present paper we analyse three-dimensional matter collapse
 with initial conditions (ICs) that are quasi-spherical, i.e., the ICs amount, to the zeroth order in a perturbation parameter $\epsilon$, to a spherical problem, and to first order in $\epsilon$ a geometrical perturbation that breaks spherical symmetry.
In the following we will not make any assumption about the form of this geometrical perturbation (that can depend on all three space coordinates), 
and thus leave it as a free function.

For the given scenario,
perturbed initial conditions can be formulated in terms of a superposition of two contributions to the initial gravitational potential, 
the first being the spherical ('top-hat') part and the second one a small asymmetric perturbation. Specifically we write for the Hessian of the initial gravitational potential
\be \label{ICs}
  \varphi_{,ij}^{\rm (init)} = \delta_{ij} \frac A 3 +  \epsilon \phi^{\rm (init)}_{,ij} \,,
\ee
where $A$ is a positive constant function,  $\epsilon$ a small perturbation parameter and  $\phi^{\rm (init)}$ an arbitrary function of all three space variables.
The case $A <0$, which resembles the evolution of a void, is not treated in the present paper and will be investigated elsewhere (see~\citealt{Sahni:1995rr,NadkarniGhosh:2010th} for low-order approximations).

Taking into account the slaving conditions~\eqref{slaving}, we have the following relation between the gradients of the initial velocity and gravitational potential,
\be \label{initialvelocity}
 v_{i,j}^{\rm (init)} \stackrel{!}{=} - \varphi_{,ij}^{\rm (init)} = - \delta_{ij} \frac A 3  -  \epsilon \phi^{\rm (init)}_{,ij} \,.
\ee

\subsection{The Lagrangian perturbation \textit{Ansatz}}\label{sec:solAnsatz}

We employ a perturbation method in which the solutions to the Lagrangian equations are
expanded in powers of $\epsilon$. Specifically, we impose for the 
particle trajectories
\be
 \fett{x}(\fett{q},a) = \fett{q} + \fett{\xi}^{(0)}(\fett{q},a) + \epsilon \fett{\xi}^{(1)}(\fett{q},a) + \epsilon^2 \fett{\xi}^{(2)}(\fett{q},a) + \ldots ,
\ee
where $\fett{\xi}^{(n)}$ is the $n$th coefficient in the $\epsilon$-expansion for the 
displacement $\fett{x}-\fett{q}$. Evidently, for the zeroth order in the $\epsilon$ expansion, we have the spherical problem;
we call this {\it the unperturbed problem}.
For the unperturbed problem the tensor of displacement gradients must be isotropic, thus 
\be
  \xi_{i,j}^{(0)} = \delta_{ij} S \,,
\ee
where $S$ is a time-dependent unknown and $\delta_{ij}$ the Kronecker delta. 
In the present paper we only expand to first order in~$\epsilon$, henceforth we write  $\fett{\xi}^{(1)}(\fett{q},a) = \fett{\xi}(\fett{q},a)$ and neglect all higher-order terms.
We thus impose for the Jacobian matrix
\be \label{ansatz_jacmatrix}
 x_{i,j} = \delta_{ij}(1+S) + \epsilon \xi_{i,j}  + O(\epsilon^2)\,,
\ee
and for its determinant, the Jacobian of the {\it perturbed problem},  
\be \label{pertJac}
 J_\epsilon = J^{(0)} + \epsilon (1+S)^2  \xi_{l,l}  + O(\epsilon^2) \,,
\ee
where $J^{(0)} = (1+S)^3$ is the unperturbed Jacobian. 

As evident from the Lagrangian mass conservation~\eqref{lagmass}, the density blows up when the Jacobian vanishes. Thus, the vanishing of the Jacobian can be used as an indicator that matter has collapsed to high-density objects. 
In more mathematical terms, the first vanishing of the Jacobian marks the instance of first shell-crossing, i.e., the time where particle trajectories begin to intersect and the single-stream description breaks down.

In the following section we solve for the particle trajectories to zeroth order in $\epsilon$. 
Then, in section~\ref{sec:pert}, we include the asymmetrical perturbation in the problem and show that the particle trajectories are time analytic and thus representable by a convergent time-Taylor series until the final stage of the non-linear collapse.

\section{The unperturbed problem (E\lowercase{d}S universe)}\label{sec:unpert}

\subsection{Spherical collapse: Equations and solutions to order \texorpdfstring{$\fett{\epsilon}^{\fett{0}}$}{epsilon}}\label{spheric}

Exact analytical solutions to the spherical problem are known in the literature, but are always investigated by considering a Friedmann toy model. Here we 
approach the problem in a more flexible environment, 
namely by solving the fluid equations.
We note that a very similar Lagrangian approach to ours, however restricted to low-order approximations, has been applied by \citet{Munshi:1994zb} and \citet{Yoshisato:1997eb}. Here we go, formally, to all orders, which allows us to proof mathematical convergence of the perturbation series.

Plugging the {\it Ansatz} to order $\epsilon^0$ in equations~\eqref{eq:Cauchy} gives a trivial identity, whereas for
 equation~\eqref{eq:scalarLag} we obtain
\begin{align}  \label{evomain}
  (1+S)^2 \R S = \frac 3 2  \left[  S  + S^2 + \frac{S^3}{3} \right] \,.
\end{align}
To solve this equation, we seek a nested {\it Ansatz} for the unperturbed displacement
in terms of a power series around $a=0$,
\be \label{ansatzunpert}
  S(a) = - \sum_{n=1}^\infty \sigma_n (Aa)^n  \,,
\ee
where $\sigma_n$ are numerical coefficients to be determined.  Note the minus sign and the powers of $A$ in our {\it Ansatz}, due to convention. 

Using this {\it Ansatz} and applying the slaving conditions~\eqref{slaving}, we obtain  directly the first-order solution  $\sigma_1 = 1/3$.
To solve for the higher-order Taylor coefficients, we plug~\eqref{ansatzunpert}
into~\eqref{evomain}; identifying the powers in $Aa$ we then get ($n>1$)
\begin{align}
  &\left( n + \frac 3 2 \right) \left( n-1 \right) \sigma_n = 
    \sum_{p+q=n} \left( 2 q^2 + q -\frac 3 2  \right) \sigma_p \sigma_q \nonumber \\
   &\qquad - \sum_{k+l+m=n} \left( m^2 +\frac m 2 - \frac 1 2  \right)  \sigma_k \sigma_l  \sigma_m \,.
\end{align}
After symmetrizing the terms on the r.h.s.\ and division of the coefficients in front of the l.h.s., we obtain
the recursion relations for the coefficients of the unperturbed displacement \mbox{($n \geq 1$)}
\begin{align} \label{rec:unpert}
 \sigma_n &= \frac{1}{3} \delta_{n1}  
   + \sum_{q<n}  \frac{q^2 + (n-q)^2 - (3-n)/2  }{(n + 3/2) ( n-1)} \sigma_q \sigma_{n-q} \nonumber \\
   & -  \sum_{k+l+m=n} \frac{ k^2 + l^2 + m^2 - (3-n)/2}{3(n + 3/2) ( n-1)}  
   \sigma_k \sigma_l \sigma_m \,.
\end{align}
The first Taylor coefficients are
\begin{align}
\begin{aligned} 
 \sigma_1 &= \frac 1 3  \,, \quad  \sigma_2 =  \frac{1}{21}\,, \quad  \hspace{0.1cm}
  \sigma_3 = \frac{23}{1701} \,, \quad \sigma_4 = \frac{1894}{392931} \,.
\end{aligned}
\end{align}
The first three coefficients $\sigma_1 - \sigma_3$ can be found in \citet{Munshi:1994zb}.  \citet{Sahni:1995rr} derived the Taylor coefficients up to order $n=5$, however for a void spherical top-hat and thus some of their coefficients have different signs.
\citet{Wagner:2015gva} determined $\sigma_1 - \sigma_5$ within the separate universe approach (see their eq.\,B.15); the match
of these coefficients could reveal an interesting relationship between the separate universe and Lagrangian-coordinates approaches, and should be investigated further.  

To our knowledge, the recursion relation~\eqref{rec:unpert} has not been reported before in the literature.
 None the less, as we show briefly now, this result can be set in direct context to standard calculations of the density in the spherical collapse model.
Indeed, by plugging our results  for the displacement into the definition of the density at the Lagrangian position, $\delta=1/J^{(0)}-1$, with $J^{(0)}=(1+S)^3$, and Taylor expanding we find 
\begin{align} \label{eq:deltaSPT}
  \delta = \sum_{n=1}^\infty \frac{\nu_n}{n!} (A a)^n \,, 
\end{align}
with the first non-vanishing coefficients
\begin{align}
  \nu_1 = 1 \,, \quad \hspace{0.2cm} \nu_2 = \frac{34}{21} \,, \quad  \hspace{0.1cm}
  \nu_3 = \frac{682}{189} \,, \quad \hspace{0.15cm}
  \nu_4 = \frac{446440}{43659} \,.
\end{align}
These coefficients agree with the ones obtained from \cite{Bernardeau:1992zw}, however 
we emphasize that our approach is different to theirs; 
obtaining these coefficients in the present
paper is little more than a check that our methodology can be directly connected to
existing works in the literature.
Furthermore, we are not aware of any literature that establishes 
the mathematical convergence of 
eq.\,\eqref{eq:deltaSPT} until collapse, although we note that there exist
explicit recursion relations for the density [cf.\ \citealt{Bernardeau:2001qr}; see also eqs.\,\eqref{recsSPT}].

\subsection{Spherical case: Convergence until collapse}\label{sphericsol}

After having found recursive solutions for the Taylor series of the unperturbed displacement, it is natural to ask the question: is 
 \be
  S(a) = - \sum_{n=1}^\infty \sigma_n (Aa)^n 
\ee
a convergent series and thus defines an exact solution until shell-crossing?

\begin{figure}
\includegraphics[width=0.477\textwidth]{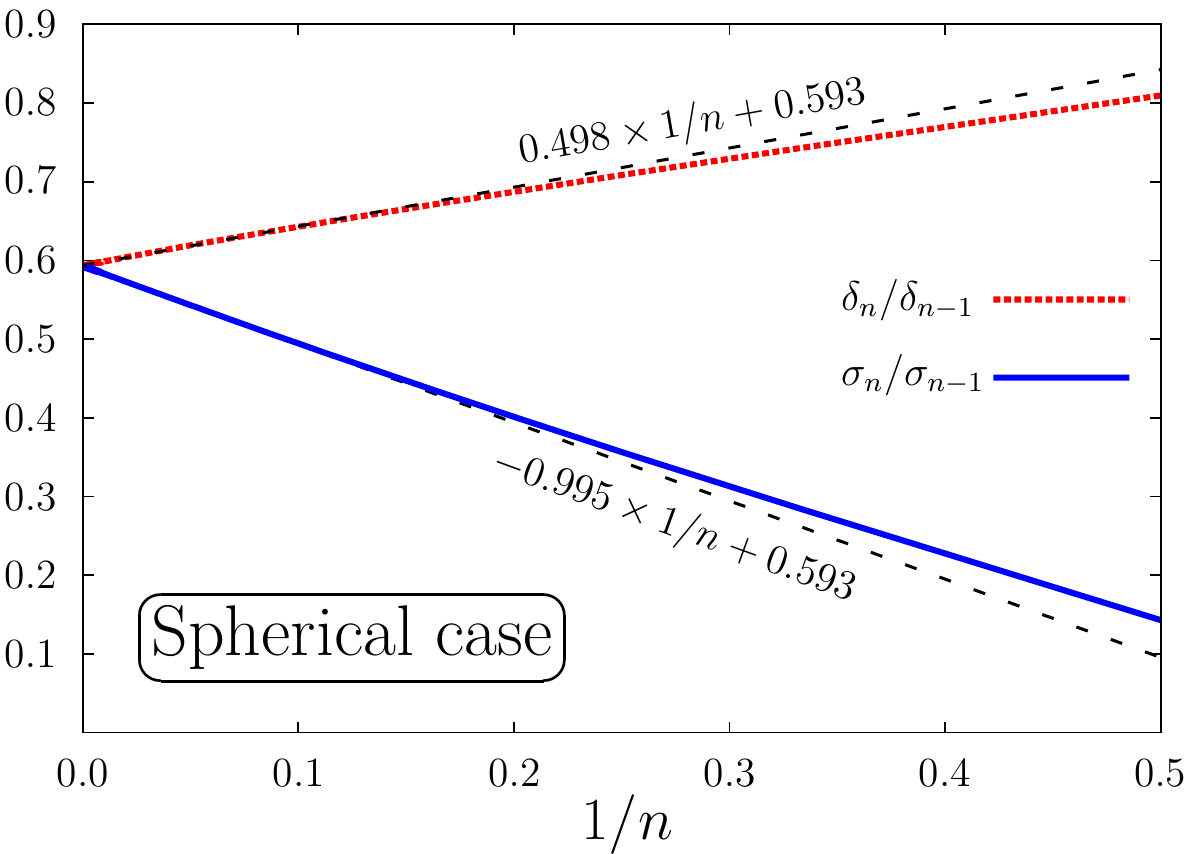} 
\vskip-0.1cm\caption{Domb--Sykes plot for the unperturbed collapse. 
Shown are ratios of the 
Taylor coefficients  $\delta_n = \nu_n/n!$ for the density contrast (red dotted line) and of the displacement coefficients  $\sigma_n$ (blue solid line).
Both ratios of subsequent coefficients approach $0.593$
 for $n \to \infty$ (obtained by a linear fit for $10 \lesssim n < 800$; dashed lines) and thus mark
the radius of convergence of the time series at $|Aa_\star^{(0)}| = 1/0.593 = 1.686$ in the complex time disc.
Formally evaluating the unperturbed Jacobian, $J^{(0)}$, at the real time value of $Aa_\star^{(0)} = 1.686$,
it is seen that the radius of convergence is limited by the instance of first shell-crossing, where  $J^{(0)}$ vanishes and the density becomes infinite.
}
\label{fig:ratios}
\end{figure}

To address this question, we perform the ratio test which states that the radius of convergence $R$ of the series is given by the relation
\be
  \frac 1 R = \lim_{n \to \infty} \frac{\sigma_n}{\sigma_{n-1}}
\ee
(if that limit exists),
where $\sigma_n$ are the Taylor coefficients of the unperturbed displacement field~\eqref{ansatzunpert}.
We determine the radius of convergence of the time-Taylor series of the displacement by drawing the Domb--Sykes plot \citep{DombSykes}, shown in fig.\,\ref{fig:ratios} (blue solid line). To obtain this plot we have generated Taylor coefficients for the displacement (and the density; red dotted line) up to order $n=800$; the output, though quite lengthy at large Taylor orders, can be easily obtained by employing standard computer algebra programs and by the use of our recursion relations. As evident from fig.\,\ref{fig:ratios}, for sufficiently large Taylor orders ($n>10$), both ratios of Taylor coefficients settle into a linear behaviour. By linearly extrapolating the ratios, shown as dashed lines, we obtain the value $0.593$ at the intersection of $1/n=0$, from which we conclude that the radius of convergence is, for both the displacement and density, given by $Aa_\star^{(0)} = 1/0.593 = 1.686$.

The radius of convergence of a series is determined by the nearest singularity in the complex disc of its argument. For a time-Taylor series, complex time singularities generally restrict the time for which the time series does converge; however, in most cases it is possible to extend the time of validity by employing suitable analytic continuation techniques, see \citet{NadkarniGhosh:2010th} for related discussions within the spherical collapse model, and \cite{CL2016} for highly related techniques for incompressible Euler flow.
In the present case, at least for the displacement, it is a priori not ruled if the first singularity occurs for real times. 
To clarify the nature of the singularity at the disc of convergence, we perform numerically a Cauchy convergence test for the 
series coefficients $S_n \equiv - \sigma_n (Aa)^n$ of the displacement ($S = \sum_n S_n$), at the critical vicinity of $Aa_\star^{(0)} = 1.686$.

Evaluating the Cauchy test to orders as large as $n=1000$, we find that indeed the series converges absolutely until the real time value of $Aa_\star^{(0)} = 1.686$. 
As a direct consequence, we can solve for the displacement from initial time $a =0$ until $a = a_\star^{(0)} = 1.686/A$, where convergence is guaranteed. Evaluating the unperturbed Jacobian $J^{(0)} \equiv (1+S)^3$ at this maximal time, one finds
\be
 J^{(0)}(a_\star^{(0)}) = 0 \,,
\ee
and thus, as expected, $a_\star^{(0)}$ marks the time of first shell-crossing / matter collapse in the unperturbed case.
Therefore, in Lagrangian coordinates, we can solve for the particle trajectories all the way to the collapse, and for that only a single time step is required.

\begin{figure}
\vskip0.05cm
\includegraphics[width=0.477\textwidth]{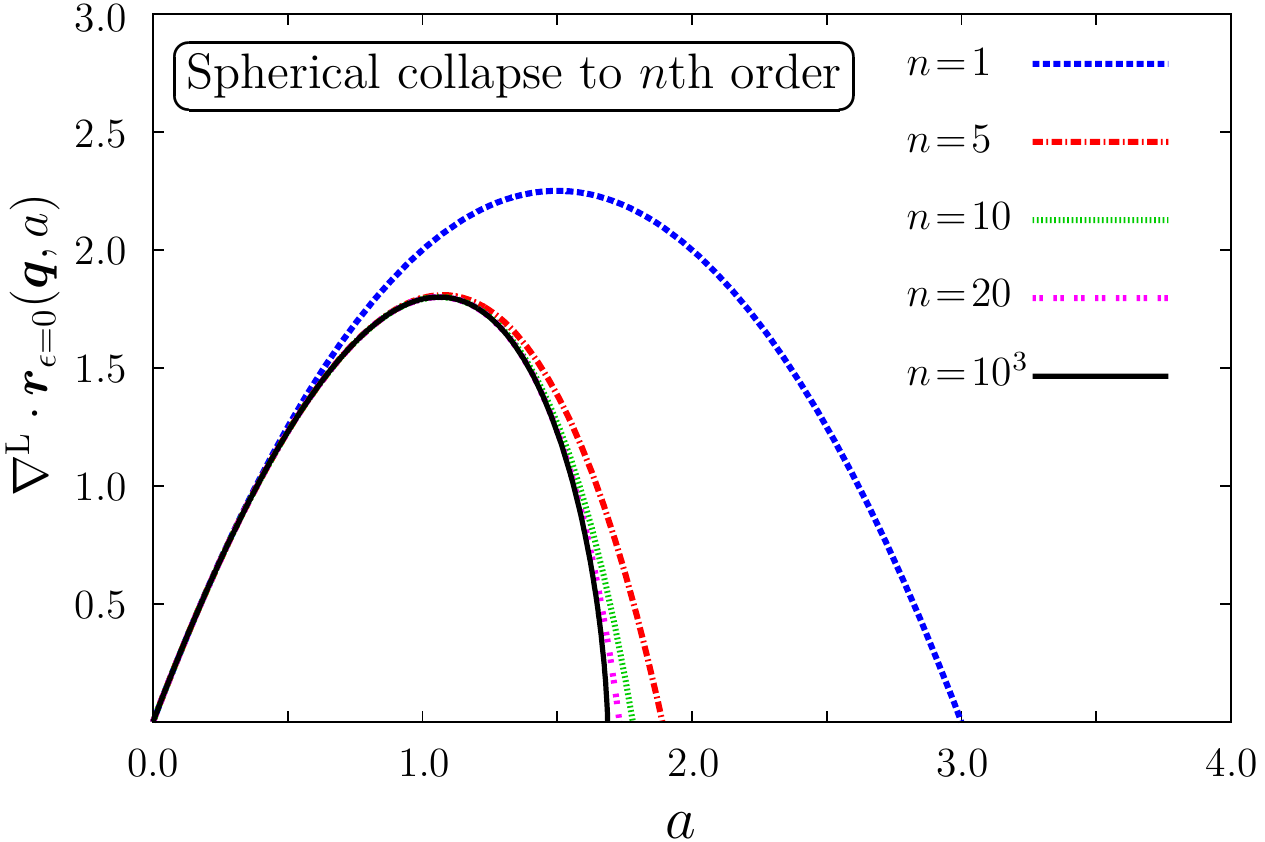}
\vskip-0.1cm\caption{Unperturbed turnaround and collapse, here for $A=1$. Shown is the divergence of the non-comoving particle trajectory 
approximated to $n$th order in the time-Taylor expansion, where $n=1,5,10,20$, and $1000$ (top to bottom lines). Beyond order $n \gtrsim 100$, the trajectory does not change visibly, and converges to the results of the spherical collapse model. 
We remark that we use the cosmic scale factor as time variable, not cosmic time; this explains why the above plot appears to be deformed in comparison to the standard spherical collapse results (see e.g.\ fig.\,2 in \citealt{LoVerde:2014rxa}).
}
\label{fig:unpert_turnaround}
\end{figure}

Also shown in fig.\,\ref{fig:ratios} is the ratio of Taylor coefficients of the density contrast 
\be
 \delta = \sum_{n=1}^\infty \frac{\nu_n}{n!} (A a)^n = \sum_{n=1}^\infty \delta_n (A a)^n \,.
\ee 
The $\nu_n$ coefficients are determined by the recursion relations \citep{Bernardeau:2001qr}
\begin{align} \label{recsSPT}
  \nu_n &= \sum_{m=1}^{n-1} \begin{pmatrix} n \\ m \end{pmatrix}
     \mu_m \frac{(2n+1) \nu_{n-m} + 2 \mu_{n-m}/3}{(2n+3)(n-1)} \, , \\
 \mu_n &= \sum_{m=1}^{n-1} \begin{pmatrix} n \\ m \end{pmatrix}
    \mu_m  \frac{3 \nu_{n-m} + 2n \mu_{n-m}/3}{(2n+3)(n-1)} \, , 
\end{align}
themselves being the result of a spherical average of the perturbative Eulerian density and velocity divergence, respectively. 
Similarly as above, we draw the Domb--Sykes plot for the Taylor series of the density. As evident from the red [dotted] line in fig.\,\ref{fig:ratios}, the radius of convergence for the density is identical with the one for the displacement. This coincidence is because of the vanishing  of the convective term in the Euler equation (second term on the l.h.s.\ in~\eqref{eq:Euler}), due to spherical symmetry. For non-isotropical ICs, the convective term does generally not vanish, and as a consequence we expect the radius of convergence of the Eulerian density to be smaller than the radius of convergence of the displacement (cf.\ \citealt{Rampf:2015mza}).

There is an even more striking argument why the Lagrangian-coordinates approach is superior compared to the Eulerian one, even for the simplistic case of perfect sphericity.
At the instance of shell-crossing, the density is indeed a real singularity; approaching it by Eulerian means, and in a controlled way is impossible. Even slightly before the time of shell-crossing, when the density is not yet infinity but very large, very high orders in the Taylor series of the density are required to resolve the matter collapse in its final stages.
In the Lagrangian approach, by contrast, the displacement is the only dynamical variable, and behaves fairly smoothly at shell-crossing [cf.\ upper panel of fig.\,\ref{fig:multiplotJacobian}, showing $J^{(0)}=(1+S)^3$ which controls the inverse of the density].

\subsection{Further results on the unperturbed problem}

\begin{figure}
\center
\includegraphics[width=0.465\textwidth]{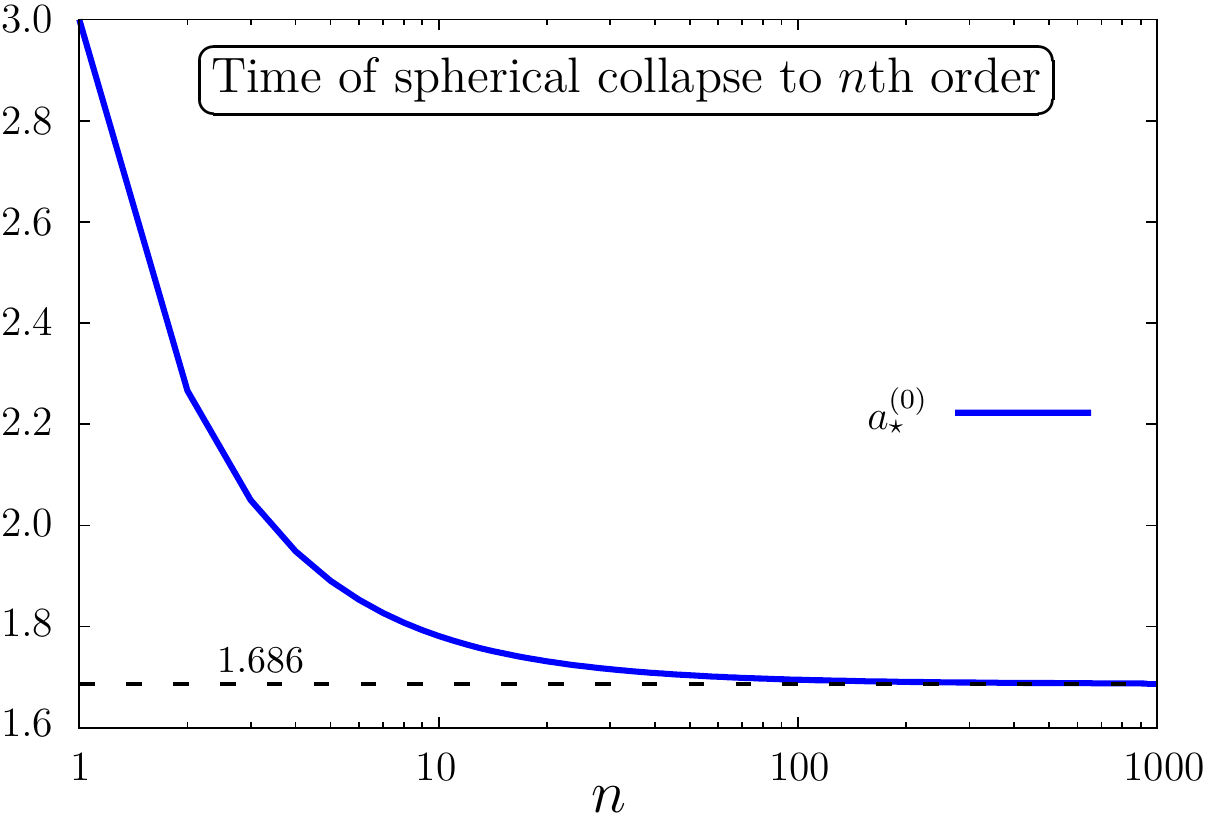}
\vskip-0.01cm\caption{Time of unperturbed shell-crossing, here for $A=1$. Shown is the 
$n$th-order approximation of $a_\star^{(0)}$, as the  result of solving numerically for the vanishing of the unperturbed Jacobian. Evidently, the numerical approximation asymptotes to the exact value of 1.686, shown as dashed line.
}
\label{fig:tstar0}
\end{figure}

Taking the trace of the unperturbed Jacobian matrix $x_{i,j} = \mbox{$\delta_{ij}(1+S)$} + O(\epsilon)$, 
and multiplying it by the scale factor, we arrive at
the divergence of the non-comoving particle trajectory,
\be \label{eq:nctrajectory}
  \nabla^{\rm L} \cdot \fett{r}_{\epsilon=0}(\fett{q},a) 
  \equiv 3 a ( 1 + S) \,,
\ee
shown  in fig.\,\eqref{fig:unpert_turnaround} for several levels of accuracies in the time-Taylor expansion. Evidently, low-order approximations for the trajectory perform poorly, none the less even the first-order solution, in Lagrangian space, predicts the existence of a turnaround and a collapse. This should be contrasted to Eulerian perturbation theory which, at first order, does not predict a collapse.

Going to higher orders in the time-Taylor coefficients, the trajectory converges quite quickly to a stable answer. More in detail, beyond order $n\gtrsim 100$  and for times $0\leq Aa \leq 1.6$, the corrections to the exact solution of the trajectory are less than $0.07$\,\textperthousand, with the largest deviation at the latest time. Higher orders are only required when evaluating the final stages of the collapse. Indeed, resolving this highly non-linear regime, we find that the time-Taylor series must be truncated up to order $n=950$ to obtain better than $0.6$\,\% precision for $1.6 \leq Aa \leq 1.685$.

We note that the time of unperturbed shell-crossing could also be  obtained by numerically evaluating $J^{(0)}(a_\star^{(0)})=0$ to a given order $n$, and the approximative results for $a_\star^{(0)}$ are shown in fig.\,\eqref{fig:tstar0}. The accuracy for obtaining $a_\star^{(0)}$  numerically gets increasingly better at large Taylor orders, and we find that,
at $n=1000$, the time of unperturbed shell-crossing can be obtained to an accuracy 
of better than $0.05$\% w.rt.\ the exact prediction of the spherical collapse model, which is (see e.g.\ \citealt{Peebles1980})
\be
 A a_{\star \rm scm}^{(0)} = \frac 3 5 \left( \frac{3 \upi}{2} \right)^{2/3} \simeq 1.68647 \,.
\ee
In our framework, however, a much better prediction can be obtained by using the extrapolation method that leads to our fig.\,\ref{fig:ratios}. Indeed, by drawing the Domb--Sykes plot for the Taylor coefficients to order $n=800$, we obtain a prediction for $A a_{\star}^{(0)}$ which is two orders of magnitude better than the numerical method as employed in fig.\,\ref{fig:tstar0}.

Numerically evaluating for the time of 'turn around', by contrast, delivers very accurate predictions already at fairly low Taylor orders. Specifically, the time $Aa_{\rm ta}$ of unperturbed turnaround is achieved when the first time derivative of the non-comoving trajectory~\eqref{eq:nctrajectory} vanishes. The standard parametric solution in the spherical collapse model gives (see e.g.\ \citealt{1994ApJ...431..486B})
\be
  Aa_{\rm ta, scm} = \frac{3}{20} (6\upi)^{2/3} \simeq 1.06241 \,,
\ee
whereas numerically evaluating for $Aa_{\rm ta}$, with our methods and to fixed time-Taylor orders $n=10,20$ and $100$, yields a precision  of better than $10^{-4}, 10^{-5}$ and $10^{-15}$, respectively.

\section{The perturbed problem (E\lowercase{d}S universe)}\label{sec:pert}

\subsection{Quasi-spherical case: equations and solutions to order \texorpdfstring{$\fett{\epsilon}^{\fett{1}}$}{epsilon}}\label{quasi}

Collecting all terms $O(\epsilon)$, we obtain from equations~\eqref{eq:scalarLag} and~\eqref{eq:Cauchy} respectively
\begin{subequations} \label{eq:main}
\begin{align}
 &  \left(1+S \right)^3 \,\R  \xi_{l,l} = \frac 3 2 \xi_{l,l}  \left[ 1 + S +S^2 + \frac{S^3}{3} \right]  \,, \label{eq:pert}  \\
 & \left( 1+S \right) \varepsilon_{ijk} \dot \xi_{j,k} = \dot S\, \varepsilon_{ijk} \xi_{j,k} \,,
\end{align}
\end{subequations}
where we remind the reader that the overdot stands for a time derivative w.r.t.\ the scale factor $a$.
The last equation dictates that no transverse displacement is generated during the evolution, and thus, to order $\epsilon$, the perturbed displacement is purely potential. Therefore, the perturbed displacement is entirely described by its divergence part, which we define as
\be
  \nabla \cdot \fett{\xi} \equiv Q  \,.
\ee
Furthermore, since the perturbed equation~\eqref{eq:pert} is autonomous in the space variables, and  because the only spatial scale is given by the perturbed initial conditions, it follows that we can write $Q$ in separable form,
\be
   Q (\fett{q},a) = - \chi(a) \,A^{-1}  \Delta^{\rm (init)}(\fett{q}) \,.
\ee
The latter space-dependent function is fully determined by the perturbed initial conditions~\eqref{ICs}, supplemented with the initial constraint $\dot\chi(a=0) =A$, it is
\be
  \Delta^{\rm (init)}(\fett{q})  =  \nabla^2 \phi^{\rm (init)} \,. 
\ee
 Thus, the space dependence of the perturbed solution is already imprinted in the initial conditions of the perturbed problem, and we only need to solve for the time dependence, given by $\chi$ which is subject to the time differential equation
\be
  \left(1+S \right)^3 \,\R \chi = \frac 3 2 \chi  \left[ 1 + S +S^2 + \frac{S^3}{3} \right]  \,. \label{eq:pert2}
\ee 
This is our evolution equation for the perturbed problem that we solve by imposing the time-Taylor series {\it Ansatz}
\be \label{ansatzchi}
  \chi = \sum_{n=1}^\infty \chi_{n} (Aa)^n \,.
\ee 
The first-order solution, determined by the slaving condition, is simply $\chi_1 = 1$.
To get the solutions for the time-Taylor coefficients for $n>1$, we plug the {\it Ansatz} into the evolution equation~\eqref{eq:pert2}. 
Matching the time-Taylor coefficients at fixed order, we get for $n>1$
\begin{align} \label{recpert}
  &\left( n + \frac 3 2 \right) \left( n-1 \right) \chi_n = 
     3\sum_{p+q=n} \left( q^2 +\frac q 2  - \frac 1 2  \right) \sigma_p \chi_q \nonumber \\
   &\qquad - 3\sum_{p+q+r=n} \left( r^2 +\frac r 2 - \frac 1 2 \right) \sigma_p \sigma_q \chi_r  \nonumber \\
   &\qquad + \sum_{p+q+r+s=n} \left( s^2 +\frac s 2 - \frac 1 2 \right) \sigma_p \sigma_q \sigma_r \chi_s \,,
\end{align}
which, after symmetrization, yields a recursion relation for $\chi_n$.
Here we will not show the explicit recursion relations as derived from~\eqref{recpert}, mainly because the involved
symmetrization of the terms on its r.h.s.\  becomes fairly cluttered (see eq.\,\eqref{recLambdaPert} and in there $\lambda=0$).
Instead and much simpler, we find that $\chi_n$ are entirely determined by the following recursion relation ($n\geq1$)
\be \label{solchi}
   \chi_n = 3 n \,\sigma_n \,,
\ee
where the $\sigma_n$'s are given by their own recursion relation~\eqref{rec:unpert}. 
We prove the validity of this trivial recursion relation in appendix~\ref{app:simple}, 
which appears to be only valid for an EdS universe (and thus does not apply for a $\Lambda$CDM Universe, see section~\ref{sec:LCDM}).
For future reference, we report here the first Taylor coefficients for the perturbed displacement,
\begin{align} \label{chi1-4}
\begin{aligned} 
 \chi_1 &= 1  \,, \quad \,  \chi_2 =  \frac 2 7 \,, \quad \,
  \chi_3 = \frac{23}{189} \,, \quad\, \chi_4 = \frac{7576}{130977} \,, 
\end{aligned}
\end{align}
which, to our knowledge, have not yet been reported in the literature.
Furthermore, because the time-Taylor coefficients of the perturbed displacement are intrinsically related to the coefficients of the unperturbed displacement, it is trivial to determine the radius of convergence of the time-Taylor series $\chi = \sum_n \chi_n (Aa)^n$. Indeed, performing the ratio test for its coefficients, we find
\be
  \frac 1 R = \lim_{n\to \infty} \frac{\chi_n}{\chi_{n-1}} 
    = \lim_{n\to \infty} \frac{3 n \sigma_n}{3 (n-1) \sigma_{n-1}} 
   =  \lim_{n\to \infty} \frac{\sigma_n}{\sigma_{n-1}} \,,
\ee
and thus, the radius of convergence of the series representation of $\chi$ is identical with the one for $\sigma$, namely $R = A a_\star^{(0)}$.
Note however, that in the perturbed case, the absolute value of that radius of convergence is {\it not} identical with the time of perturbed shell-crossing, the latter denoted with $a_\star$. This is simply because the total trajectory is a superposition of the unperturbed trajectory and the perturbed displacement, 
both coming with their individual validity regime.
Rather, as we show in the following section, the time of perturbed collapse occurs generically earlier than in the unperturbed case, i.e., $a_\star \leq a_\star^{(0)}$.
This allows us to solve exactly for the perturbed particle trajectory (including the unperturbed part) before and at the instance of perturbed shell-crossing.

It is also interesting to plug the r.h.s.\ of equation~\eqref{solchi} into the {\it Ansatz} for $\chi$, which reveals a novel relationship between the perturbed and unperturbed displacement and the fluid velocity,
\begin{align}
  \chi &= \sum_{n=1}^\infty \chi_n (Aa)^n  = 3 \sum_{n=1}^\infty  n \sigma_n (Aa)^n  \equiv  - 3a \dot S \,,
\end{align}
where $\dot S = \upartial_a^{\rm L} S = - \upartial_a^{\rm L} \sum_n \sigma_n (Aa)^n$ is the fluid velocity of the unperturbed theory.

Summing up, from the above results we obtain, to order $\epsilon$, respectively the Jacobian matrix and the Jacobian
\begin{subequations} \label{mainresult}
\begin{align}
 & x_{i,j} = \delta_{ij} \left( 1+ S \right)
    - \epsilon \frac {\chi}{A}  \nabla_{\rm L}^{-2} \upartial_i^{\rm L} \upartial_j^{\rm L} \Delta^{\rm (init)}  \,,  \label{xij} \\
& J_\epsilon = (1+S)^3  - \epsilon   (1+S)^2 \frac {\chi}{A}  \Delta^{\rm (init)} \,, \label{fullJ}
\end{align}
\end{subequations}
with $S= \sum_n \sigma_n (Aa)^n$ and $\chi = -3 a \dot S$, where the $\sigma_n$'s are given by eq.\,\eqref{rec:unpert}.
We remark again, that $S$
is {\it independent of the chosen initial conditions for} $\Delta^{\rm (init)}= \nabla^2 \phi^{\rm (init)}$, and thus the above results hold for an arbitrary choice of initial conditions.

Equations~\eqref{xij} and~\eqref{fullJ} constitute the main technical results of this paper, which we will explore in the following two sections.

\subsection{The time of perturbed shell-crossing/matter collapse}\label{sec:time}

In the absence of any perturbations, spherical collapse occurs at $Aa_{\star}^{(0)} = 1.686$.  Since the leading-order correction to the displacement is linear in $\epsilon$,
it is expected that the time of matter collapse receives a correction linear in $\epsilon$ as well. Our solution {\it Ansatz} for the time of perturbed collapse is therefore
\be
  A a_\star = A a_{\star}^{(0)} + \epsilon C \,,
\ee
where $C$ is a constant which can only depend on the space coordinates. 
Perturbed shell-crossing occurs at the time $a_\star$ for which the Jacobian vanishes,
\be \label{analysisJac}
 J_\epsilon(a_\star) = \left[1+S(a_\star) \right]^2 \left( 1 +S(a_\star) + \epsilon  \frac {3 a_\star \dot S(a_\star)}{A}   \Delta^{\rm (init)} \right) =0 .
\ee
Evidently, {\it this Jacobian vanishes also at the time of unperturbed shell-crossing}, $a_\star^{(0)}$, for which the square bracketed term vanishes.
However,
as we argue above, in the perturbed scenario shell-crossing could be shifted to earlier times, in which case we expect the deciding contribution in~\eqref{analysisJac} coming from the round bracketed term. 
Assuming that $a_\star < a_\star^{(0)}$ for the moment, and to leading order in $\epsilon$, we can ignore the overall factor of $[1+S(a_\star)]^2$ in the last equation, and  thus are left with 
\be \label{here}
  1 + S(a_\star) + \epsilon  \frac {3 a_\star \dot S(a_\star)}{A}  \Delta^{\rm (init)}  = 0   \,.
\ee
Now, since $S(a_\star) =  S(a_\star^{(0)}) + \epsilon C \dot S(a_\star^{(0)})/A +  O(\epsilon^2)$, and because of $1+S(a_\star^{(0)}) =0$, it is straightforward to find 
from equation~\eqref{here} that
$C = - 3a_\star^{(0)} \Delta^{\rm (init)}$. Thus, the time of perturbed shell-crossing is, to order $\epsilon$, and for times  $a_\star < a_\star^{(0)}$,
\be \label{tstar}
   A a_\star = A a_{\star}^{(0)} ( 1  -  3 \epsilon  \Delta^{\rm (init)}/A) \,, 
\ee
with $A a_{\star}^{(0)} = 1.686$ and $\Delta^{\rm (init)} = \nabla^2 \phi^{\rm (init)}$.
Since $\Delta^{\rm (init)}$  can take generally also positive values, we thus conclude that if $\Delta^{\rm (init)}>0$ locally, then indeed perturbed shell-crossing occurs earlier than in the unperturbed case.
Stated in another way, an initially overdense region will collapse earlier, if the perturbation $\Delta^{\rm (init)}$ amplifies the initial overdensity. 

\begin{figure}
\includegraphics[width=0.477\textwidth]{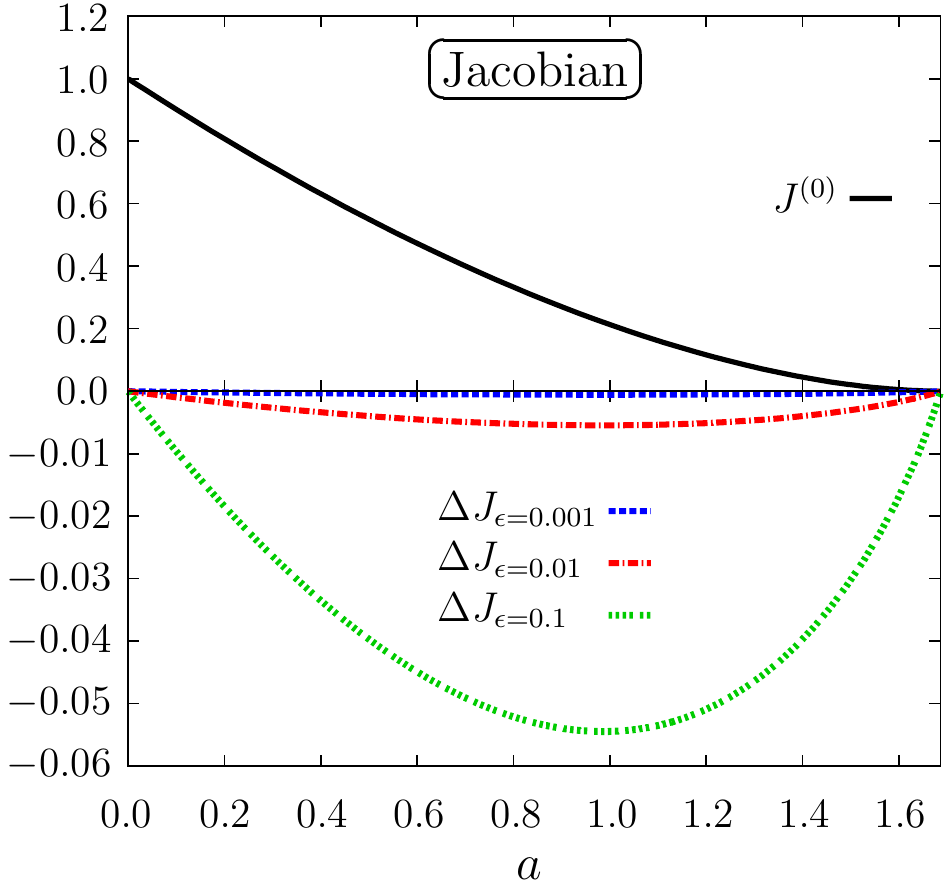}
\vskip-0.1cm\caption{{\it Upper panel:} The unperturbed Jacobian $J^{(0)}$ as a function of the $a$-time. {\it Lower panel:} Difference between the perturbed and unperturbed Jacobian, i.e.,  $\Delta J_\epsilon = J_\epsilon - J^{(0)}$ for several values of the perturbation \mbox{parameter $\epsilon$.} For these plots we have set $A=1$ and $\Delta^{\rm (init)}=1$.
}
\label{fig:multiplotJacobian}
\end{figure}

What about locations $\fett{q} = \fett{Q}$ for which $\Delta^{\rm (init)}(\fett{Q}) \leq 0$? Will the time of perturbed shell-crossing be delayed w.r.t.\ the unperturbed case? The answer to this question is no, since what matters physically is the {\it first} vanishing of the Jacobian~\eqref{analysisJac}, which is guaranteed to happen, at the latest, at the time of unperturbed shell-crossing (for which the square bracketed term in~\eqref{analysisJac} vanishes). 

Summing up, to leading order in $\epsilon$, and for $\Delta^{\rm (init)} > 0$ perturbed shell-crossing occurs as instructed by equation~\eqref{tstar}, but for $\Delta^{\rm (init)} \leq 0$, the time of perturbed shell-crossing is identical with the time of unperturbed shell-crossing.
 
To our knowledge,
the qualitative observation that perturbations to the spherical collapse can lead to a decrease of the time of collapse has been made the first time by \citet{Monaco:1997cq}, who investigated the ellipsoidal collapse up to third order in Lagrangian perturbation theory.
There, the time of collapse has been determined by solving numerically for the first vanishing of the Jacobian. This solution technique, 
when restricted to the perturbed problem (and thus not to arbitrary large deformations for which we can make no positive statements on the convergence), however, converges very slowly, as very high perturbation orders are required to accurately resolve the Jacobian at the final stages of the collapse (cf.\ our fig.\,\ref{fig:tstar0}).
The presented results, by contrast, though restricted to sufficiently small departures from spherical symmetry, are exact results -- represented in terms of fully converging Taylor series -- results that can be determined to arbitrary high accuracy.

\subsection{Further results on the perturbed problem}\label{sec:pertfurther}

In fig.\,\ref{fig:multiplotJacobian} we show the unperturbed Jacobian as well as the difference $\Delta J_\epsilon = J_\epsilon - J^{(0)}$,\, for several values of the perturbation parameter~$\epsilon$. For simplicity we have set $A=1$ and $\Delta^{\rm (init)}$.
Noticeable from that figure is that the effect of the perturbation yields 
the largest deviation from the unperturbed Jacobian at the time of turnaround,
the latter defined by the  maximum value of the divergence of the non-comoving particle trajectory.
This is most easily seen on the following fig.\,\ref{fig:full_turnaround}, where we plot the physical particle trajectories, for the same values of $\epsilon$ as in the last figure.

\begin{figure}
\vskip0.1cm
\includegraphics[width=0.477\textwidth]{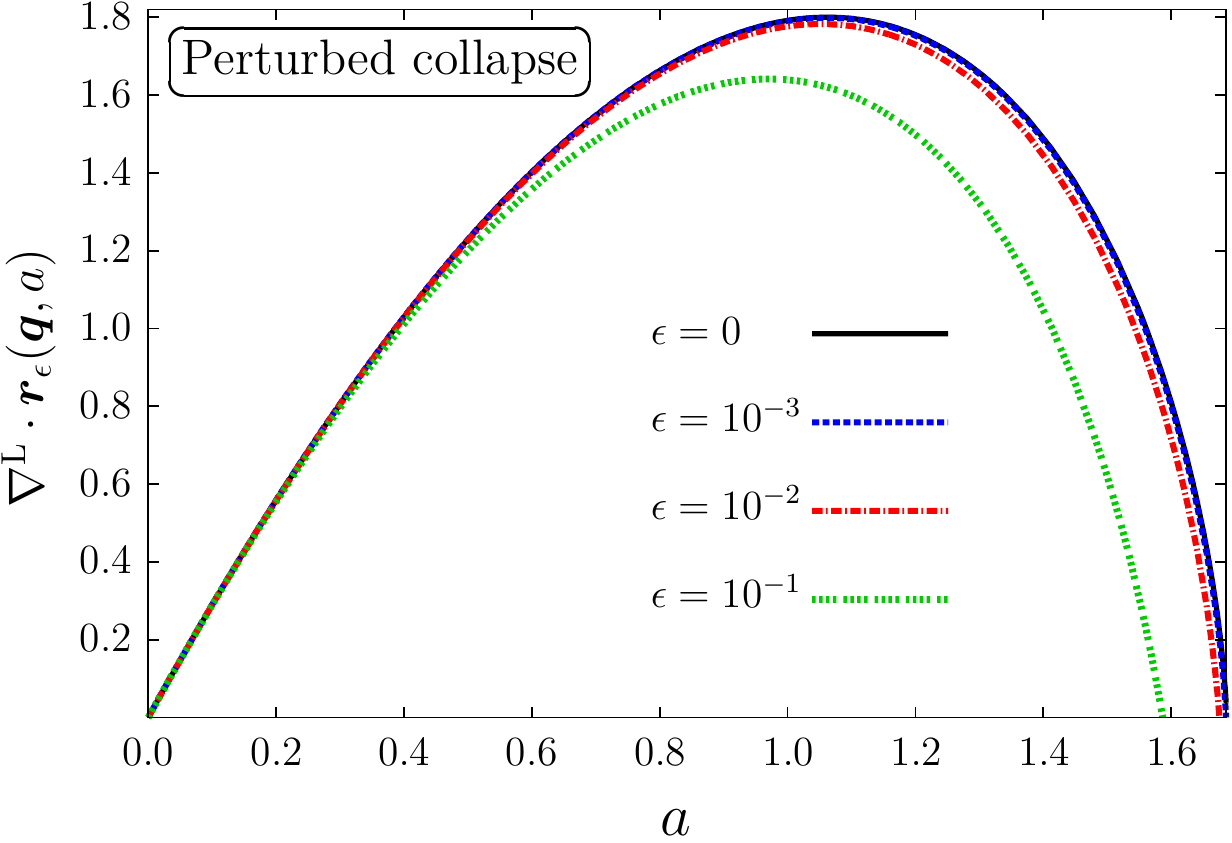}
\vskip-0.1cm\caption{The perturbed non-comoving (i.e, physical) particle trajectory,
for $A=1$ and $\Delta^{\rm (init)} =1$. Shown are several choices of values of the perturbation parameter, where $\epsilon=0$ refers to the unperturbed trajectory (black line). The case $\epsilon =10^{-3}$ (blue line) is almost exactly overlapping with the unperturbed case. Generally, the larger the value of $\epsilon$, the earlier is the time of  turnaround and collapse.
}
\label{fig:full_turnaround}
\end{figure}

Let us discuss the consequences for the density in the perturbed case.
Since the fully non-linear density becomes infinite at the collapse, the non-linear
density is not a useful quantity to determine.
The linearized density, however, is still a well behaved quantity even at the collapse, and can be thus useful (e.g., the linear density is a central input quantity in halo formation models).
In the present paper, we perform essentially a double expansion scheme, one being an expansion around $a=0$ and the other being a perturbative departure from exact sphericity, the latter parametrized \mbox{with $\epsilon$.}
Thus, in the present double expansion, a linearization involves taking the leading-order terms in $\epsilon$ as well as in $a$.

Performing the just described linearization of the density $\delta =1/J-1$, 
we obtain the linearized density
\be 
  \delta_{\rm lin}(a) = Aa \left( 1 + \epsilon \Delta^{\rm (init)}/A\right) \,.
\ee
Evaluating the linear density contrast at the critical vicinity of perturbed shell-crossing $a_\star$, which for  $\Delta^{\rm (init)} >0$ is given by eq.\,\eqref{tstar}, and otherwise is identical with $a_\star^{(0)}$, we then find  
\begin{align} \label{deltacrit}
  \delta_{\rm lin}(a_\star) &=  1.686 \left( 1 - c\, \epsilon |\Delta^{\rm (init)}|/A\right)  \,,
\end{align}
where
\be
   c = \left\{ \begin{matrix} 1,  \quad \text{for} \quad \Delta^{\rm (init)} < 0 \\  2,    \quad \text{for} \quad \Delta^{\rm (init)} >0 \end{matrix} \right. \,.
\ee
Thus, irrespective of the sign of the perturbation to the spherical collapse, the threshold for the linear density contrast at collapse is decreased w.r.t.\ to the unperturbed problem.

\section{Generalization to a \texorpdfstring{$\fett{\Lambda}$CDM}{LCDM} Universe}\label{sec:LCDM}

Our results can be easily generalized to more realistic cosmologies, such as to the spatially flat
$\Lambda$CDM Universe that includes, apart from CDM, also a cosmological constant $\sim \Lambda$.
To employ our developed tools, we first need the Lagrangian evolution equations for a dark matter fluid in $\Lambda$CDM. 
These equations and a thorough derivation are given by~\citet{Rampf:2015mza}.
Although the $\Lambda$CDM Universe is nowadays the commonly accepted cosmological model and thus the literature of $\Lambda$CDM vast, 
the present formulation is only little known, hence 
for the sake of clarity we briefly summarize its derivation.

We begin with the fluid equations in physical (i.e., non-comoving) coordinates which are \citep{Peebles1980}
\begin{subequations}
\begin{align}
& \partial_t {\fett{U}} + \left( {\fett{U}} \cdot \nabla_r \right) {\fett{U}} =  - \nabla_r  \phi_{\rm g} \,, \\
 \label{eq:massmn} & \partial_t  \varrho +\nabla_r \cdot \left(  \varrho {\fett{U}}\right) =0 \,, \\
  \label{eq:lambdapoisson} &\nabla_r^2  \phi_{\rm g} = 4\uppi G \varrho - 3\Lambda \,,
\end{align} 
\end{subequations}
with $\fett{r}$ the proper space coordinate, $t$ the cosmic time, $\fett{U}$ the physical velocity (including the Hubble term), $\varrho$ the fluid density, and the cosmological constant given by $3 \Lambda$.
Using the decomposition
\be
  \fett{r} = a(t)\, \fett{x} \,, \quad 
  \varrho = \bar \rho(t) [1 +\delta ] \,, \quad
  \fett{U} = H(t)\, {\fett{r}} + a \fett{u} 
\ee
in the above fluid equations, where $H(t) = (\partial_t a)/a$ is the usual Hubble parameter, we obtain for the purely time-dependent background part the well-known Friedmann equation,
\be
  H^2 = a^{-3} + \Lambda \,,
\ee
where for notational simplicity we have absorbed some of the standard coefficients into $\Lambda$. 
For the mass and velocity fluctuations $\delta$ and $\fett{u} = \partial \fett{x}/ \partial t$,  the fluid equations become 
\begin{subequations}
 \begin{align}
 \label{eq:EulerL}
& \partial_t {\fett{u}} + \left( {\fett{u}} \cdot \nabla_x \right) {\fett{u}} = - 2 H  \fett{u}  - \frac{3}{{2a}^2} \nabla_x  \varphi_{\rm g} \,, \\
 \label{eq:massL}
& \partial_t \delta +\nabla_x \cdot \left[ (1+\delta) \fett{u} \right] =0 \,, \\
 \label{eq:PoisL}
& \nabla_x^2  \varphi_{\rm g} = \frac \delta a \,.
 \end{align}
\end{subequations}
In these fluid equations, although being indeed valid for $\Lambda$CDM, there is no explicit appearance of the cosmological constant $3\Lambda$, since the background part has been substracted out. Instead, the $\Lambda$-dependence is imprinted in the time evolution of the cosmic scale factor $a(t)$ (and $H$), itself determined by the Friedmann equations.

Perturbative solutions to arbitrary high order for for the fluid equations are most easily obtained by changing from cosmic time $t$ to the $a$-time. 
This task is straightforward by noting that the time-derivatives are related via \mbox{$\partial_t = (\partial a / \partial t)\, \partial_a$,} and using the first Friedmann equation to get an expression for $(\partial a / \partial t)$.
Then one obtains the so-called peculiar fluid equations in the $a$-time formulation \citep{Rampf:2015mza}
\begin{subequations} \label{fluidLambdaE}
\begin{align}
  \label{eq:Eulera}
&\!\left(  1+ \Lambda a^3 \right) \left[  \partial_a \fett{v} + \left( \fett{v} \cdot \nabla_x \right) \fett{v} \right] = -
\frac{3}{2a} \left( \fett{v} + \nabla_x \varphi_{\rm g} \right) - 3 \Lambda a^2 \fett{v} \,, \\
   \label{eq:massa}
&\partial_a \delta +\nabla_x \cdot \left[ \left( 1+ \delta \right) \fett{v} \right] = 0  \,, \\
   \label{eq:Poisa}
& \nabla_x^2 \varphi_{\rm g} = \frac \delta a \,,
\end{align}
\end{subequations}
with $\fett{u} = (\partial_t a) \,\fett{v}$. One note is in order.
Apart from the scale-factor time, there is another convenient time variable for $\Lambda$CDM calculations, namely the linear growing mode for $\Lambda$CDM which is usually called $D$ (cf.\ \citealt{Hamilton:2000tk}). However, as argued by~\citet{Rampf:2015mza}, the resulting perturbative expressions take its simplest form when expressed in the $a$-time, hence our choice for the scale-factor time.

Finally, we transform equations~\eqref{fluidLambdaE} to Lagrangian space, which yields
\begin{subequations} \label{LeqsLCDM}
 \begin{align}
  & \varepsilon_{ikl} \varepsilon_{jmn}  x_{k,m}  x_{l,n} \,\R^{(\Lambda)}  x_{i,j}
    = 3 \left( J-1 \right) \,, \\
  &\varepsilon_{ijk} \, \dot x_{l,j} x_{l,k} = 0  \,,
 \end{align}
\end{subequations}
with notations and conventions as in the previous sections, except with a new temporal operator which we define with
\be \label{RLambda}
 \R^{(\Lambda)} \equiv [ a^2 ( 1+\Lambda a^3 ) (\partial^{\rm L}_{a})^2 + 3 \Lambda a^4 \partial^{\rm L}_{a} + (3a/2)\partial^{\rm L}_{a}] \,.
\ee
Observe that, apart from the slightly updated temporal operator which has two additional terms $\sim \Lambda$, 
the Lagrangian evolution equations for $\Lambda$CDM {\it are formally identical with those for EdS} (cf.\ eqs.\,\eqref{eqs:lag}). 
This correspondence allows us to employ the same tools as in the previous sections and  to obtain all-order solutions for the displacement.

\subsection{Spherical collapse in $\Lambda$CDM}

\begin{figure}
\includegraphics[width=0.477\textwidth]{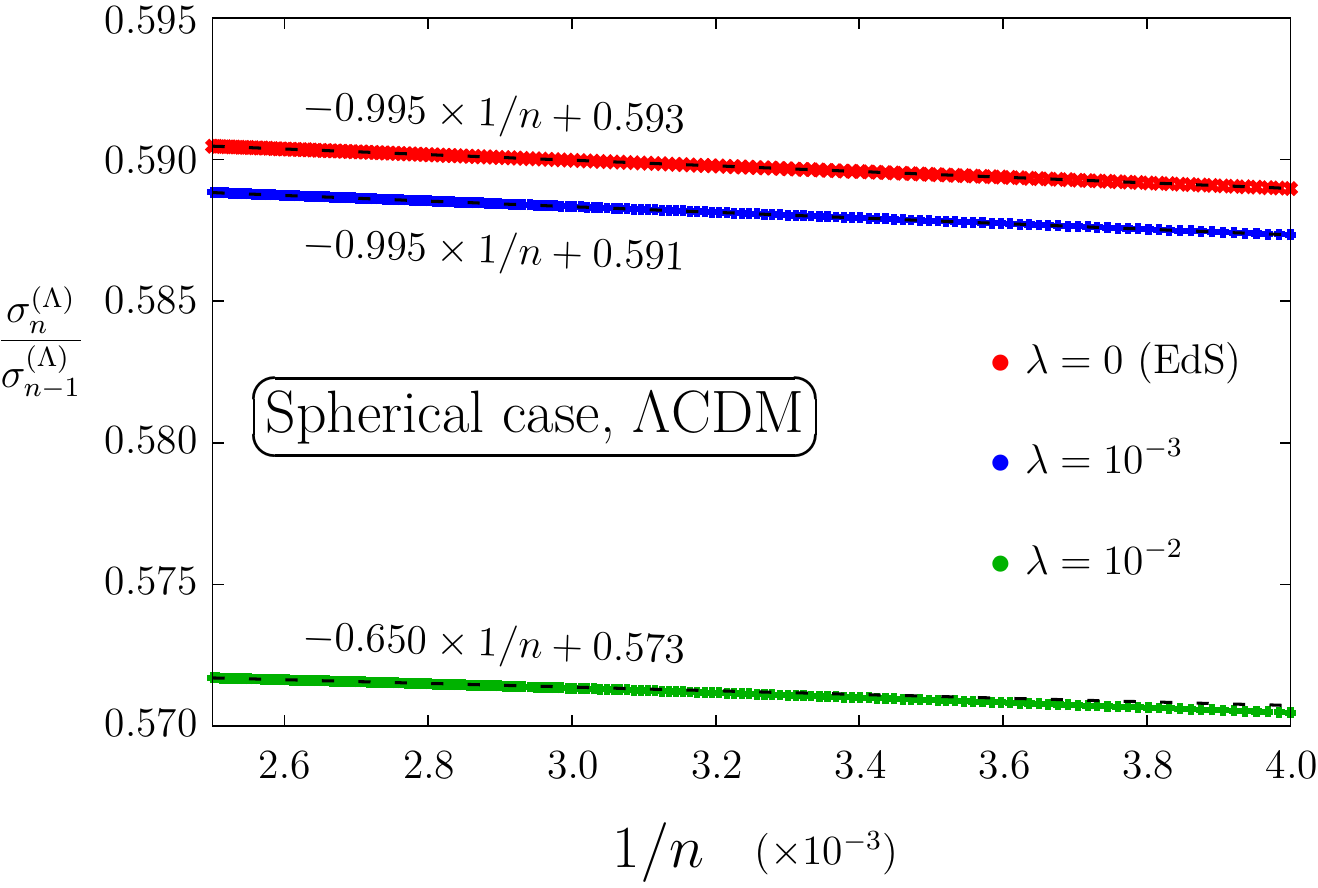} 
\vskip-0.1cm\caption{Domb--Sykes plot for the spherical collapse in $\Lambda$CDM, zoomed in at large Taylor orders $250 \leq n \leq 400$. 
Shown are subsequent ratios of the 
Taylor coefficients of the displacement coefficients  $\sigma_n^{(\Lambda)}/\sigma_{n-1}^{(\Lambda)}$, 
for several choices of the free parameter $\lambda = \Lambda/A^3$.
The (mostly) overlapping dashed lines are the result of linear extrapolations, obtained by linear fits for $350 \lesssim n \leq 400$.
The linear extrapolations for $n \to \infty$ indicate that the inverse 
of the radius of convergence is an increasing function 
\mbox{of $\lambda$.} For the radii of convergence we find 
$|A a_\star|  = 1.686, 1.692, 1.745$ for \mbox{$\lambda = 0, 10^{-3}, 10^{-2}$.}
Formally evaluating the physical trajectory (see also fig.\,\ref{fig:phystrajLCDM})
at this critical time value,
it is seen that the radius of convergence is limited by the collapse.
}
\label{fig:ratiosLCDM}
\end{figure}

Let us investigate the spherical collapse in $\Lambda$CDM for which we need to solve 
for the $\Lambda$CDM displacement $S_{\Lambda}$ in the equation
\be
   (1+S_{\Lambda})^2\, \R^{(\Lambda)} S_\Lambda = \frac 3 2  \left[  S_{\Lambda}  + S_\Lambda^2 + \frac{S_\Lambda^3}{3} \right] \,,
\ee
with the operator $\R^{(\Lambda)}$ given in~\eqref{RLambda}.
We impose for the displacement 
\be
  S_{\Lambda} (a) = - \sum_{n=1}^\infty \sigma_n^{(\Lambda)} (Aa)^n  \,. \label{SL}
\ee
Plugging this {\it Ansatz} into eq.\,\eqref{LeqsLCDM} and identifying the powers in
$(Aa)^n$, we obtain the following all-order recursion relation for the Taylor coefficients \mbox{($n \geq 1$)}
\begin{align} \label{SLsol}
  \sigma_n^{(\Lambda)} &= \frac 1 3 \delta_{n1}  - \frac{\Lambda}{A^3} \frac{n-3}{n+3/2} \sigma_{n-3} \nonumber \\
    &+ \sum_{q<n} \frac{q^2 + (n-q)^2 - (3-n)/2}{(n+3/2)(n-1)} \sigma_q \sigma_{n-q} \nonumber \\
     &+ \frac{\Lambda}{A^3}  \sum_{q<n} \frac{q^2+ (n-q)^2 -4n + 6}{(n+3/2)(n-1)} \sigma_q \sigma_{n-q-3} \nonumber \\
   &- \sum_{k+l+m=n} \frac{k^2+l^2+m^2 - (3-n)/2}{3(n + 3/2) (n-1)} \sigma_k \sigma_l \sigma_m \nonumber \\
 &- \frac{\Lambda}{A^3}  \sum_{k+l+m=n} \frac{k^2+l^2+m^2 -4n+9}{3(n + 3/2) (n-1)} \sigma_k \sigma_l \sigma_{m-3} \,,
\end{align}
where we demand that $\sigma_k^{(\Lambda)} = 0$ if $k \leq 0$.
Comparing this recursion relation against the one for EdS, given in eq.\,\eqref{rec:unpert}, it becomes evident that in the $\Lambda$CDM case there are double as many terms involved, thanks to $\Lambda$. (As can be easily checked, setting $\Lambda =0$ returns the EdS result.)

Furthermore, because of the given structure of the $\Lambda$CDM recursion relation,
Taylor coefficients involving $\Lambda$ only appear for $n \geq 4$.
Explicitly, the first Taylor coefficients are
\begin{align}
\begin{aligned} 
 &\sigma_1^{(\Lambda)} =  1/ 3  \,, \quad  \sigma_2^{(\Lambda)} =  1/21\,, \qquad   \sigma_3^{(\Lambda)} = 23/1701 \,, \\
  &\sigma_4^{(\Lambda)} = \frac{1894}{392931} - \frac{2 \Lambda}{33 A^3}\,, \hspace{0.3cm}
  \sigma_5^{(\Lambda)} =  \frac{3293}{1702701} -  \frac{179 \Lambda}{6006 A^3} \,.
\end{aligned}
\end{align}
Furthermore, for orders $n>6$,
higher-order Taylor coefficients are generally populated by powers of $(\Lambda/A^3)^m$, with $m \in \mathbb{N}$ and
$1 \leq m \leq n -6$.
The explicit appearance of $\Lambda$ and of the top-hat amplitude $A$ in the recursion
relations renders the collapse problem inherently scale-dependent, 
an expected phenomenon when departing from EdS.

\begin{figure}
\vskip0.05cm
\includegraphics[width=0.477\textwidth]{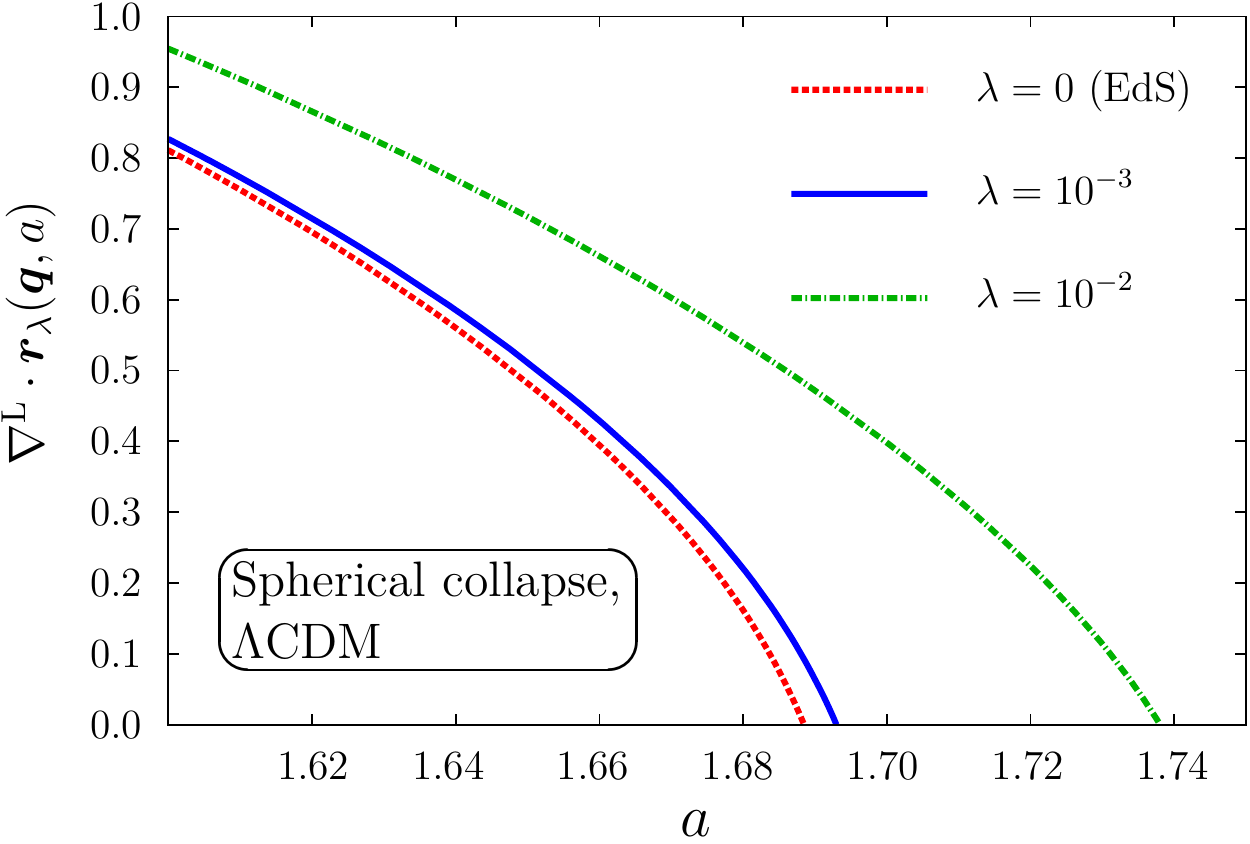}
\vskip-0.1cm\caption{Spherical collapse in $\Lambda$CDM, here for $A=1$. 
Shown is the divergence of the non-comoving particle trajectory at the final stages of the collapse, for the same values of  $\lambda$ 
as in the previous figure. 
To obtain the $\Lambda$CDM trajectories we have evaluated the displacement 
up to \mbox{order $n=400$.} 
The collapse times, for which 
$\nabla^{\rm L} \cdot \fett{r}_{\lambda}=0$,
according to the present figure agree well with the
linear extrapolations from the Domb-Sykes method (fig.\,\ref{fig:ratiosLCDM}).
}
\label{fig:phystrajLCDM}
\end{figure}

Having obtained the all-order Taylor-series representation for the spherical collapse 
in $\Lambda$CDM, let us investigate the convergence properties of its series. 
Clearly, the introduction of a non-zero $\Lambda$, although only visibly present at large Taylor orders, should affect the radius of convergence, since it is indeed the large 
Taylor orders that determine the convergence properties (or non-convergence) of a series.
We perform the ratio test to obtain the radius of convergence~$R^{(\Lambda)}$ of the Taylor series~\eqref{SL} in $\Lambda$CDM, with
\be
  \frac{1}{R^{(\Lambda)}} = \lim_{n \to \infty} \frac{\sigma_n^{(\Lambda)}}{\sigma_{n-1}^{(\Lambda)}} \,.
\ee 
In fig.\,\ref{fig:ratiosLCDM} we show the numerical results for several choices of 
the free parameter $\lambda \equiv \Lambda/A^3 = 0, 10^{-3}, 10^{-2}$. 
To obtain this figure, we have determined the Taylor coefficients up to order $n=400$. 
The linear extrapolations for each parameter choice indicate that convergent solutions of the series~\eqref{SL} are obtained for the maximal time values of
\begin{align} \label{Astar0LCDM}
  |A a_{\star \Lambda}^{(0)}|  = 
    \begin{cases}
      1/0.593 = 1.686   & \text{for } \lambda=0 \\  
      1/0.591 = 1.692  & \text{for } \lambda=10^{-3} \\
      1/0.573 = 1.745   & \text{for } \lambda=10^{-2}  
    \end{cases} \,.
%
\end{align}
Thus, the radius of convergence increases with larger $\lambda$. 
Formally evaluating the divergence of the unperturbed physical trajectories ($\fett{r} = a \fett{x}$) at that critical time value 
\be \label{phystrajLCDM}
  \nabla^{\rm L} \cdot \fett{r}_{\lambda}(\fett{q},a_{\star \Lambda}^{(0)}) = 3 a\, \left[ 1 + S_{\Lambda}(a_{\star \Lambda}^{(0)}) \right] \to 0 \,,
\ee
it becomes evident that $|A a_{\star \Lambda}^{(0)}|$ marks the time of unperturbed collapse (for which $\nabla^{\rm L} \cdot \fett{r}_\lambda=0$). 

This can also be seen in~fig.\,\ref{fig:phystrajLCDM} where we show the physical trajectory~\eqref{phystrajLCDM} in $\Lambda$CDM. For larger values of $\lambda$ (and for fixed $A$) the cosmological constant opposes the collapse stronger, and thus the collapse becomes delayed.
From that figure we can also read off the collapse times, which, for sufficiently small $\lambda$, agree excellently with those as predicted by the Domb-Sykes method. For the trajectory with the largest shown value of $\lambda = 10^{-2}$, we observe a slight mismatch of collapse times of the order of $0.4\%$, indicating that higher Taylor orders in the displacement~\eqref{SL} are required for larger values of $\lambda$ whilst keeping the accuracy goal.
Actually, the fact that the displacement is not yet fully converged at order $n=400$
for $\lambda = 10^{-2}$ can also be seen in the previous figure~\ref{fig:ratiosLCDM}: 
evidently, the ratios of subsequent Taylor coefficients have not yet settled into a linear behaviour even at orders $n \approx 300$.

For larger values of $\lambda$ than $\sim 10^{-1}$ but for fixed $A$ the collapse will not occur at all, since the strong accelerated expansion washes the density fluctuations away (not shown). 
At increasingly larger values for $\lambda$, it is furthermore expected that the Taylor series will cease to converge, however we consider this large $\lambda$-scenario as physically less relevant (besides the seemingly close analogy with standard inflationary theory which however requires a relativistic description).

We note that the used $\lambda$ values for fig.\,\ref{fig:phystrajLCDM}  are 
only exemplary to show the effect of a non-zero $\Lambda$; realistic values for $\Lambda$ would not deliver any visible effect for the present perturbative expansion, since the expansion is performed around $a=0$, i.e., initial data for the collapse is
formally provided at $a=0$, a property of our mathematical model. 
At such early times, we are deep in the matter era and thus $\Lambda$ has little influence on the matter dynamics.
Of course, our tools can also be used for late-time initializations, and
in section~\ref{sec:disc} we provide an in-depth discussion about such avenues.

\subsection{Quasi-spherical collapse in $\Lambda$CDM}

Continuing to a perturbed collapse in $\Lambda$CDM, we require to
find a solution to the following equations at $O(\epsilon)$,
\begin{subequations} \label{eq:mainL}
\begin{align}
  & \left(1+S \right)^3 \,\R^{(\Lambda)}  \xi_{l,l} = \frac 3 2 \xi_{l,l}  \left[ 1 + S +S^2 + \frac{S^3}{3} \right]  \,, \label{eq:pertL}  \\
 &\left( 1+S \right) \varepsilon_{ijk} \dot \xi_{j,k} = \dot S\, \varepsilon_{ijk} \xi_{j,k} \,.
\end{align}
\end{subequations}
As before, our strategy is to find solutions for the perturbed Jacobian matrix of the 
form $ x_{i,j} = \delta_{ij}(1+S_\Lambda + \epsilon \xi_{i,j}^{(\Lambda)})$, where the all-order solution for $S_\Lambda$ is given in eqs.\,\eqref{SL}--\eqref{SLsol}.
Because of exactly the same arguments as given in section~\ref{quasi}, the only non-zero contribution to the perturbed displacement comes from its divergence part
which we set to 
$\nabla \cdot \fett{\xi}^{(\Lambda)} \equiv Q_\Lambda = - \chi_\Lambda(a) \,A^{-1}  \Delta^{\rm (init)}(\fett{q})$. To solve for the time-dependence of the unknown, we seek solutions of the form
\be
  \chi_\Lambda = \sum_{n=1}^\infty \chi_{n}^{(\Lambda)} (Aa)^n \,.
\ee
Repeating  similar calculations as before, we then find the all-order solution ($n\geq 1$)
\begin{align} \label{recLambdaPert}
  & \chi_n^{(\Lambda)} =  \delta_{n1} - \lambda \frac{n-3}{n + 3/2} \chi_{n-3}^{(\Lambda)} \nonumber \\
  &+   3  \sum_{p+q=n}  \frac{q^2 + q/2 - 1/2}{\left( n + 3/2 \right) \left( n-1 \right)} \sigma_p^{(\Lambda)} \chi_{q}^{(\Lambda)} \nonumber \\ 
 & + 3  \lambda \sum_{p+q=n} \frac{q^2  - 4q  +3}{\left( n + 3/2 \right) \left( n-1 \right)} \sigma_p^{(\Lambda)} \chi_{q-3}^{(\Lambda)}   \nonumber \\
   &- 3 \sum_{p+q+r=n} \frac{r^2+r/2-1/2}{\left( n + 3/2 \right) \left( n-1 \right)} \sigma_p^{(\Lambda)} \sigma_q^{(\Lambda)} \chi_r^{(\Lambda)}  \nonumber \\
  & - 3 \lambda \sum_{p+q+r=n} \frac{r^2-4r+3}{\left( n + 3/2 \right) \left( n-1 \right)} \sigma_p^{(\Lambda)} \sigma_q^{(\Lambda)} \chi_{r-3}^{(\Lambda)} \nonumber \\
   & + \!\!\! \sum_{p+q+r+s=n} \!\!\!\!\! \Big\{ \frac{s^2+s/2-1/2}{\left( n + 3/2 \right) \left( n-1 \right)} \sigma_p^{(\Lambda)} \sigma_q^{(\Lambda)} \sigma_r^{(\Lambda)} \chi_s^{(\Lambda)} \nonumber \\
 & + \lambda \frac{s^2-4s+3}{\left( n + 3/2 \right) \left( n-1 \right)} \sigma_p^{(\Lambda)} \sigma_q^{(\Lambda)} \sigma_r^{(\Lambda)} \chi_{s-3}^{(\Lambda)}   \Big\} \,,
\end{align}
where  $\lambda = \Lambda/A^3$, and the terms on the r.h.s.\ still have to be symmetrized over all possible permutations. The first few Taylor coefficients are
\begin{align}
\begin{aligned} 
 &\chi_1^{(\Lambda)} = 1  \,, \qquad \,  \chi_2^{(\Lambda)} =  \frac 2 7 \,, \qquad \,
  \chi_3^{(\Lambda)} = \frac{23}{189} \,,  \\
  &\chi_4^{(\Lambda)} = \frac{7576}{130977} - \frac{2\lambda}{11}\,, \quad
   \chi_5^{(\Lambda)} = \frac{16465}{567567} - \frac{102 \lambda}{1001} \,.
\end{aligned}
\end{align}
We note that the above recursion relation is vastly different from the one in the EdS case (eq.\,\eqref{solchi}); none the less a quick check reveals that for $\lambda =0$ 
we obtain the same Taylor coefficients as in the EdS case, as required.
The obvious reason why the present recursion relation is more cluttered is, that in the $\Lambda$CDM case there is no simple way to write the perturbed displacement in terms of the unperturbed one, as it was possible in the EdS case (for the derivation see appendix~\ref{app:simple}; in particular see eq.\,\eqref{identif}).

By using our tools we have verified that the perturbed Taylor series is also convergent in $\Lambda$CDM. Actually, no rigourous proof for the convergence is required since
we have already established convergence for the perturbed displacement in EdS, 
as well as the convergence for the unperturbed displacement in $\Lambda$CDM.
Thus, particle trajectories are time-analytic from which it follows trivially that also the perturbed displacement in $\Lambda$CDM must be representable by a convergent Taylor series.

Summing up, the Jacobian matrix for $\Lambda$CDM reads in the perturbed case
\be
  x_{i,j} = \delta_{ij} \left( 1+ S_\Lambda \right)
    - \epsilon \frac {\chi_\Lambda}{A}  \phi_{,ij}^{\rm (init)} \,,
\ee
with the Taylor coefficients for $S_\Lambda$ and $\chi_\Lambda$ given in eqs.\,\eqref{SLsol} and~\eqref{recLambdaPert} respectively. From this result both fluid variables trivially follow, i.e., the density is $\delta = 1/\det[x_{i,j}]-1$,  and the velocity
is given by $\fett{v} = \partial_a^{\rm L} \fett{x}$.

Let us close this section by translating the main results obtained in the previous sections to $\Lambda$CDM. Perturbed collapse will occur at 
the time value 
$ 
 A a_{\star \Lambda} = A a_{\star \Lambda}^{(0)} + \epsilon C_\Lambda
$,
where  $A a_{\star \Lambda}^{(0)}$ denotes the time of unperturbed/spherical collapse (see e.g.\ eq.\,\eqref{Astar0LCDM}),
and $C_\Lambda$ a space-dependent constant. Actually, the analysis that we have applied in section~\ref{sec:time} straightforwardly translates to $\Lambda$CDM; in particular
we can obtain the unknown $C_\Lambda$ by plugging the {\it Ansatz} for the perturbed time into the Jacobian, evaluated to first order in~$\epsilon$, and require its vanishing. We then find
\be \label{tstarLambda}
   A a_{\star \Lambda} = A a_{\star \Lambda}^{(0)}\, ( 1  -  3 \epsilon  \Delta^{\rm (init)}/A) \,, 
\ee
and thus, perturbed collapse in $\Lambda$CDM will occur earlier than for spherical collapse if $\Delta^{\rm (init)}>0$, and otherwise occurs at the same time as in the spherical case -- exactly for the same reasons as outlined in section~\ref{sec:time}.

\section{Summary and discussion}\label{sec:disc}

The case of exact spherical collapse is highly degenerated. 
Furthermore, even just a tiny initial inhomogeneity that is added to a spherical top-hat profile is crucial, as it decides shape and orientiation of the collapsed object. Thus, such small inhomogeneities are not at all negligible and must be incorporated in realistic models for structure formation.
By departing perturbatively from the pure spherical problem, we have shown 
that the quasi-spherical problem can be solved exactly,
and by fully analytical means, until the instance of shell-crossing. The latter denotes the first crossing of particle trajectories which results in infinite densities (in Eulerian coordinates), indicating the formation of density caustics on the one side, and the break-down of the fluid description on the other.

The methodology of the present approach is as follows.
Firstly, we employ a 3D formulation of the cosmological fluid equations in
Lagrangian coordinates 
(Eulerian coordinates should be avoided when investigating the matter collapse because of the appearance of explicit singularities).
We solve the equations for a choice of initial conditions that resemble, to the zeroth order in a small expansion parameter, a spherical top-hat profile. 
In Lagrangian coordinates, the solution of the fluid equations is represented in terms
of an infinite time-Taylor series for the displacement field, for which we report all-order recursion relations [see eq.\,\eqref{rec:unpert} for EdS, and eq.\,\eqref{SLsol} for $\Lambda$CDM].
Here it is important to note that the used time variable is not the cosmic time $t$ but the cosmic scale factor $a \sim t^{2/3}$. 
By drawing the so-called Domb-Sykes plot (see fig.\,\ref{fig:ratios})
for the time-Taylor coefficients of the Lagrangian displacement field,
we establish the mathematical convergence of the Lagrangian description until collapse.
At collapse, the Jacobian of the Lagrangian transformation, which controlls the inverse density, is exactly zero, thereby
signalling the blow-up of the density.

Then, we add an arbitrary perturbation to the top-hat profile at the level of the initial conditions. This perturbation, controlled by the dimensionless perturbation parameter $\epsilon >0$,  is allowed to have any non-trivial spatial dependence, thereby breaking exact spherical symmetry. That perturbation leads to a perturbed Lagrangian displacement field, which can be represented by an infinite time-Taylor series.
For an EdS universe, we are able to vastly simplify the resulting recursion relations for the displacement, essentially expressing the perturbed displacement coefficients in terms of the unperturbed ones [eq.\,\eqref{solchi}]. For a $\Lambda$CDM Universe, we do find explicit recursions relations~\eqref{recLambdaPert} as well, which, however, can not be written in such a compact form as in the EdS case.
Then, by formally going to all orders in the Taylor series, 
we find that the series, both for EdS and $\Lambda$CDM, converge absolutely until the instance of shell-crossing.

As a direct consequence,
we obtain the perturbed particle trajectory~\eqref{xij}, subject to the initial density $\Delta^{\rm (init)}$ of the perturbation to the spherical collapse. Investigating the time of collapse~$a_\star$, for which the Jacobian vanishes the first time, it is found that 
\be
 A a_\star = 1.686 ( 1  -  3 \epsilon  \Delta^{\rm (init)}/A)
\ee
for $\Delta^{\rm (init)}>0$, and otherwise simply $A a_\star = 1.686$ for an EdS universe. Here, $A>0$ is the initial amplitude for the spherical top-hat. For $\epsilon =0$ we are back at the spherical problem, whereas for $\epsilon>0$ and $\Delta^{\rm (init)}>0$, collapse will generically occur earlier than in the pure spherical case. 
For a $\Lambda$CDM Universe, the structure of the above formula still holds, and one only needs to replace the time value for spherical collapse, which is 1.686 for EdS, by 
its $\Lambda$CDM value (which is slightly larger than 1.686, see fig.\,\ref{fig:phystrajLCDM}).
 
The observation that perturbed collapse occurs earlier than in the spherical case has been already made in the literature for specific perturbation problems, although we are only aware of fairly qualitative statements about $a_\star$, thus no analytic formula was known; see \citet{LoVerde:2014rxa} for the spherical collapse in the presence of massive neutrinos, or \citet{Monaco:1997cq} for the ellipsoidal collapse. 
We also remark that, including the perturbation in the analysis, the time of collapse as well as the (linear) density become inherently dependent on the mass scales of the collapse problem (set by the ratio $\Delta^{\rm (init)}/A$). This observation appears to be in agreement with the numerical analysis of \citet{Sheth:1999su}. 

In the present approach the collapse criterion is set by the first vanishing of the Jacobian
\be \label{collapsecr}
  J_\epsilon = \left[1+S \right]^2 \left( 1 +S + \epsilon  \frac {3 a \dot S}{A}   \Delta^{\rm (init)} \right) \,,
\ee
 which, in the perturbed problem, is triggered by the Laplacian of the perturbed initial gravitational potential $\Delta^{\rm (init)} = \nabla^2 \phi^{\rm (init)}$.
We note that the three bracketed terms on the r.h.s.\ in eq.\,\eqref{collapsecr} {\it do not} resemble the factorization 
into contributions from the three principal axis. In particular the last term in the round brackets originates from a combination from all principal axis.
Thus, it is the {\it total source of the perturbation in all coordinate directions, and not the collapse along a single coordinate axis, that sets our collapse criterion.}
As a consequence of the decreased collapse time, we find that the critical linear density at quasi-spherical collapse is reduced [see equation~\eqref{deltacrit}], irrespective of the sign of $\Delta^{\rm (init)}$. 
Finally, we remark that in the literature there exists other collapse criteria than~\eqref{collapsecr}; another frequent collapse criteria, e.g.\ used in the context of ellipsoidal or triaxial collapse models, is associated with virialization (see e.g.\ \citealt{Sheth:1999su} and the discussion therein).

For the case of a $\Lambda$CDM Universe,
the presence of $\Lambda >0$ delays the collapse w.r.t.\ to the EdS case (for which $\Lambda=0$), since the acceleration of the Universe opposes the gravitational clustering.
Generally, the physical impact of $\Lambda$ on the matter evolution is small, especially considering that we have performed a time-Taylor expansion around $a=0$ (thus formally pushing the birth of structures to the origin of time, which is however just a property of our mathematical model).
More impact of $\Lambda$ on the matter dynamics occurs at late times (see e.g.\ \citealt{Wintergerst:2010ui}); 
of course, such late-time behaviour can also be incorporated within the present methodology, provided one follows the steps as outlined in the following.

To initialize the matter collapse at arbitrary times $a_{\rm init} > 0$ within our methodology, one may proceed as follows.
Firstly, for initializations at late times, the initial density contrast $\delta_{\rm init} \equiv \delta (a_{\rm init}, \fett{q})$ is generally non-zero which needs to be incorporated in the evolution equations.
Secondly, observe that within the Lagrangian evolution equations,  the time variable appears explicitly in the temporal operator
$\R^{(\Lambda)} = [ a^2 ( 1+\Lambda a^3 ) (\partial^{\rm L}_{a})^2 + 3 \Lambda a^4 \partial^{\rm L}_{a} + (3a/2)\partial^{\rm L}_{a}]$.
The system of equations is thus not time invariant. 
Therefore, for Taylor expansions around $a_{\rm init}$, one should time-translate this operator according to $a \to a_{\rm init} + \tilde a$, where $\tilde a$ is the new time variable. The temporal operator then becomes $\R^{(\Lambda)} \to \tilde{\mathfrak{R}}_{(a_{\rm init} + \tilde a)}^{(\Lambda)}$, with all $a$'s being replaced with $a_{\rm init}+\tilde a$ and the temporal derivatives changed according to $\partial_a^{\rm L} \to \partial_{\tilde a}^{\rm L}$.

Summing up, the Lagrangian evolution equations then become
\begin{align}
  & \varepsilon_{ikl} \varepsilon_{jmn}  x_{k,m}  x_{l,n} \,\tilde{\mathfrak{R}}_{(a_{\rm init} + \tilde a)}^{(\Lambda)}  x_{i,j}
    = 3 \left( J-1 - \delta_{\rm init} \right) \,, \\
  &\varepsilon_{ijk} \, \dot x_{l,j} x_{l,k} = 0  \,,
\end{align}
which then can be solved with a Taylor series Ansatz around $a_{\rm init}$. 
These equations, which to our knowledge have never been reported in the literature and
are valid for any types of initial conditions, can then be solved by an $\tilde a$-time Taylor series. For this, note that the initial velocity in the Lagrangian representation is simply $\fett{v}_{\rm init} = \partial_{\tilde a}^{\rm L} \fett{x}(a_{\rm init}, \fett{q})$.
It is expected that the recursion relations for initializations at $a_{\rm init}>0$ will become highly non-trivial, and thus will be investigated elsewhere.

Having found new exact analytical solutions to the fluid equations, for EdS and $\Lambda$CDM (and possibly even beyond when suitably generalized), could open a new window of applications.
For example, the analytical solutions could be compared against results from $N$-body simulations, with the aim to  optimize the $N$-body technique 
in the critical vicinity of particle crossings. We find such avenues to be in 
 close correspondence with the work of e.g.\ \citet{Hahn:2014lca,Hahn:2015sia} who
have introduced new methods with the aim to refine the $N$-body technique.

Our findings of analytical solutions to the quasi-spherical collapse delivers also accurate thresholds for the critical density at collapse, which could be used as the input to formalisms that predict the abundance, mass or shape of a given tracer, for example in the Press--Schechter formalism, excursion set, or peaks theory \citep{1974ApJ...187..425P,Bond:1990iw,PhysRevD.78.103503,Paranjape:2012jt}. When suitably adapted, our methodology could be also relevant for determining the abundance of primordial black holes (cf.\ \citealt{Kuhnel:2016exn}).

Finally,
we have seen that to leading order in $\epsilon$, the perturbed displacement for quasi-spherical collapse 
is seemingly unaffected by tidal/environmental effects. 
Indeed, at no instance in the calculations have we made use of non-local operations (such as the inverse Laplacian) which
would signal environmental dependence.
The absence of environmental effects is due to the fact that, for sufficiently small departures from sphericity where a linearization in $\epsilon$ is justified, such effects are indeed negligible. 
It would be interesting to go to second order in the perturbation parameter $\epsilon$, because this would allow the inclusion of such environmental corrections (cf.\ \citealt{Desjacques:2007zg}). 
We leave such investigations for future work.

\section*{Acknowledgements}

CR thanks Uriel Frisch and Vincent Desjacques for related discussions and comments on the manuscript, 
as well as Florian K\"uhnel and Bj\"orn Malte Sch\"afer for useful discussions.
At an initial stage of this work, CR was supported by the DFG through the SFB-Transregio TRR33 ``The Dark Universe''. 
CR acknowledges funding from the People Programme (Marie Curie Actions) of the European Union H2020 Programme under grant agreement number 795707 (COSMO-BLOW-UP).



\bibliographystyle{mnras}
\bibliography{desing} 



\appendix

\section{Simple relation between unperturbed and perturbed solutions for an EdS universe}\label{app:simple}

In section~\ref{sec:pert} we reported the finding of an exact relation between the 
Taylor coefficients of the displacement, i.e., that $ \chi_n = 3 n \,\sigma_n$.
Here we prove the validity of this simple relation by considering the evolution equation for the perturbed displacement.
We note that the below is strictly valid only for an EdS universe; in particular it does not hold for a $\Lambda$CDM Universe.

Plugging the Ansatz~\eqref{ansatz_jacmatrix} for the Jacobian matrix into the evolution
equation~\eqref{eq:scalarLag} we arrive at first order in $\epsilon$ at
\be \label{alter}
  (1+S)^2 \R \chi + 2\chi (1+S) \R S = \frac 3 2 \chi (1+S)^2 \,.
\ee
The relation $\chi_n = 3 n \,\sigma_n$ between the time-Taylor coefficients amounts to the following relation,
\be \label{identif}
 \chi = 3a \dot S \,.
\ee
Using this in equation~\eqref{alter} we find 
\be \label{alter2}
   (1+S)^2 \R (a \dot S) 
  + 2 a \dot S (1+S) \R S = \frac 3 2 a \dot S (1+S)^2 \,.
\ee
Now, we  rewrite the r.h.s.\ of the last equation in terms of a Lagrangian time derivative
\be
  \text{r.h.s.} = a  \upartial_a^{\rm L} \left\{ \frac 3 2 \left[ S + S^2 + \frac{S^3}{3} \right] \right\}
   = a \upartial_a^{\rm L} \Big\{  (1+S)^2 \R S \Big\}  \,,
\ee
where in the last step we have used equation~\eqref{evomain}.
Equating the last expression with the l.h.s.\ of~\eqref{alter2}, a few terms
are cancelling without further actions, and we are left with
\be
  \R (a \dot S) = a \upartial_a^{\rm L} (\R S) \,.
\ee
This turns out to be an identity, and thus, 
we have proven that the perturbed evolution equation~\eqref{alter} is identical with the unperturbed evolution equation, {\it provided that we make use of the identification}~\eqref{identif}.

%
%
%
%

\bsp	
\label{lastpage}
\end{document}